\documentclass[twoside,11pt,onecolumn]{IEEEtran}
\pdfoutput=1
\usepackage{amsmath,amsfonts,amssymb,graphicx,cite,url,multirow,xfrac}
\usepackage{graphicx}
\usepackage[T1]{fontenc}
\usepackage[utf8]{inputenc}
\usepackage[USenglish]{babel}
\usepackage{flushend}
\usepackage{siunitx}
\usepackage{lineno}
\usepackage{xcolor}
\usepackage[normalem]{ulem}
\usepackage{url}
\usepackage[labelfont=bf]{caption}
\newcommand{\m}[1]{\textrm{#1}} 
\newcommand{\diff}{\mathop{}\!d}

\begin{document}
	\title{A Self-powered Analog Sensor-data-logging Device based on Fowler-Nordheim Dynamical Systems}
	\author{Darshit~Mehta\footnote{$^1$Department of Biomedical Engineering, Washington University in St.~Louis, 1 Brookings Drive, St.~Louis, MO, 63130 USA}$^1$, Kenji~Aono\footnote{$^2$Department of Electrical and Systems Engineering, Washington University in St.~Louis, 1 Brookings Drive, St.~Louis, MO, 63130 USA}$^2$ and~Shantanu~Chakrabartty\footnote{$^3$Email: shantanu@wustl.edu}$^{1,2,3}$}%
	\maketitle

	\begin{abstract}
		
		Continuous, battery-free operation of sensor nodes requires ultra-low-power sensing and data-logging techniques. Here we report that by directly coupling a sensor/transducer signal into globally asymptotically stable monotonic dynamical systems based on Fowler-Nordheim quantum tunneling, one can achieve self-powered sensing at an energy budget that is currently unachievable using conventional energy harvesting methods. The proposed device uses a differential architecture to compensate for environmental variations and the device can retain sensed information for durations ranging from hours to days. With a theoretical operating energy budget less than 10 attojoules, we demonstrate that when integrated with a miniature piezoelectric transducer the proposed sensor-data-logger can measure cumulative ``action'' due to ambient mechanical acceleration without any additional external power.
	\end{abstract}

	\section*{Introduction}
	For sensing systems like IoT devices or biomedical implants that operate in resource-constrained settings, utilizing a battery may be impractical due to biocompatibility concerns, size constraints or due to technical challenges involved in replacing the battery. Self-powered sensors (SPS) can obviate the need for batteries by harvesting their operational energy directly from ambient sources, such as light~\cite{indoorLight_mathews2014gaas} or mechanical vibrations~\cite{wang2006piezoelectric}. SPS achieve this by first buffering ambient energy using standard power-conditioning techniques before activating the basic computation/sensing and sometimes telemetry functions~\cite{autonomous_torah2008self,materials_mcevoy2015}. However, when the objective is to sense and compute a simple function, like the total signal energy or a cumulative ``action'', an application specific but ultra-energy-efficient variant of SPS could be designed by combining the operational physics of signal transduction, rectification and non-volatile data storage. 
	One such SPS was reported in~\cite{huang2011asynchronous, chakrabartty2010self} where a cumulative measure of mechanical activity was sensed, computed and directly stored on floating-gate memories~\cite{liangBioCas}. Similar techniques could also be applied to other non-volatile technologies for sensing the event of interest as an equivalent change in magnetoresistance in MRAM~\cite{mram_aakerman2005toward}, change polarization in FeRAM~\cite{fram1998physics}, or change in electrical conductance in memristor-type systems~\cite{memristor_chua1971}. However, these approaches require power conditioning such as rectification or voltage-boosting to meet the activation thresholds and to initiate the non-volatile state-change. Operational limits arise due to rectification efficiency, and due to material properties that influence diode thresholds or leakage currents. Note that some energy harvesting systems report low voltage continuous operation (i.e. $<$\SI{50}{\milli\V}), however they require higher activation thresholds for initial start up conditions (e.g. $>$\SI{600}{\milli\V})~\cite{ramadass2011battery, mercier2012energy,ti_bq25505}. 
	
	\begin{figure}[ht]
		\centering
		{\includegraphics[width=0.95\columnwidth]{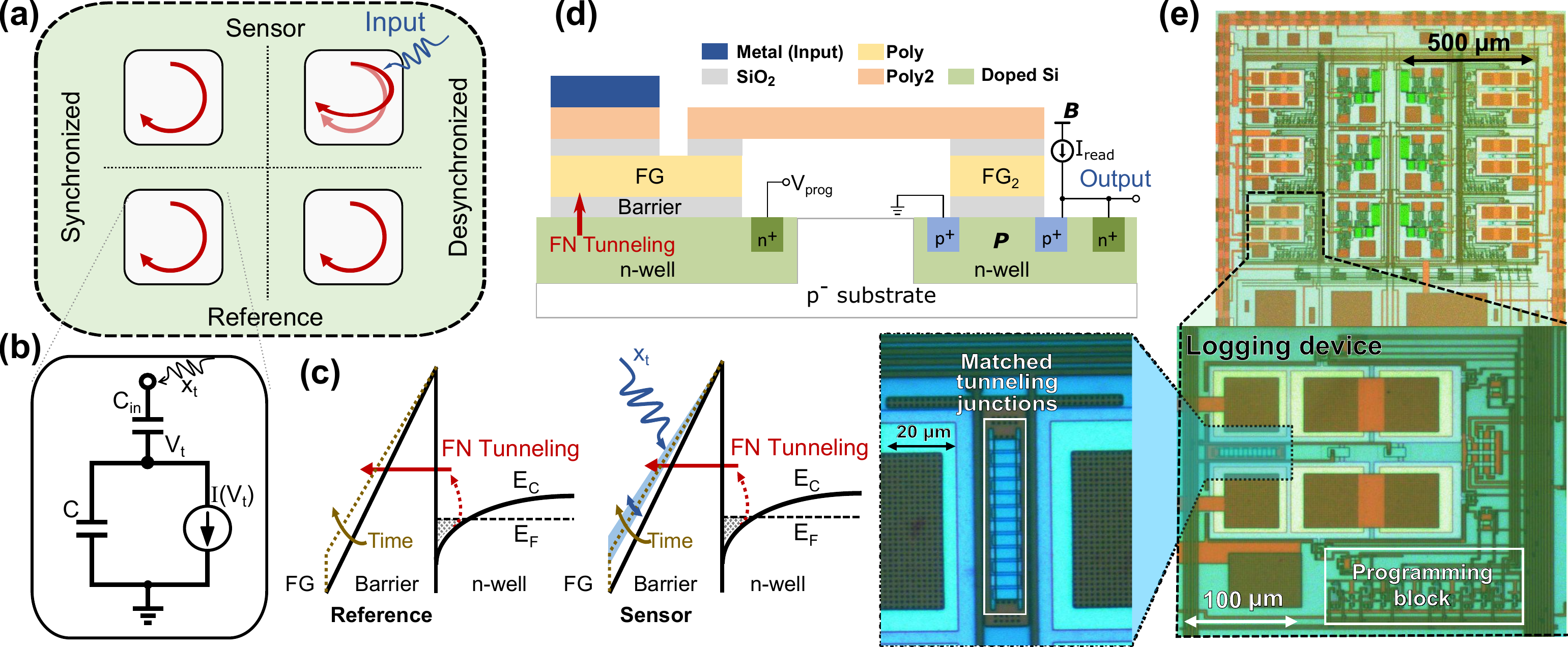}}
		\caption{\textbf{Operating principle and architecture of the proposed sensor-data-logger.} a) Principle of sensing and data-logging where the input signal leaves its trace on a pair of synchronized dynamical system through a desynchronization process. b) Equivalent circuit model of a self-powered dynamical system where the charge on a capacitor $C$ stores the dynamical state of the system and the dynamics is governed by a leakage current $I(V_t)$ and ambient stimuli $x_t$. c) Band-diagram corresponding to the tunneling junction where the electrons tunnel across the triangular energy barrier and the input signal $x_t$ modulates the barrier shape. d) Cross-section of the sensor-data-logging device showing the FN tunneling junction, the floating-gate which is coupled to a read-out transistor $P$ and a buffer $B$. e) Micrograph of the fabricated devices, with inset showing a pair of dynamical systems configured in a differential architecture.}
		\label{scheme}
	\end{figure}
	We propose a self-powered sensing system, where instead of harvesting the energy to switch between static memory states, the sensing signal is used for modulating a synchronized dynamic state. In this regard, dynamical systems, both natural and artificial, have been shown to store information in their dynamic states~\cite{informationSingleNode_appeltant2011,dynamicalInfo_dambre2012,memoryTraces_ganguli2008}. In this work, we show the feasibility of this approach for self-powered sensing and data-logging, but at chip-scale. This is illustrated in Fig.~\ref{scheme}a which shows two synchronized globally asymptotically stable (G.A.S) dynamical systems; a sensing system and a reference system. A time-varying input signal modulates the state trajectories of the sensing dynamical system leading to its desynchronization with respect to the reference dynamical system. The relative degree of desynchronization between the two systems serves as a medium for sensing and storing the cumulative effect of the input modulation. While the principle is relatively straightforward, there exists two key challenges in implementing the proposed concept at a chip-scale. First, due to self-powering requirements, the synchronized G.A.S. dynamical system can only be implemented using leakage processes driven by intrinsic thermal or quantum transport of electrons. The simplest of such a system can be modeled by an equivalent circuit shown in Fig.~\ref{scheme}b. The capacitor $C$ in the circuit models the dynamical state (denoted by the time-varying voltage $V_t$) and the time-dependent system trajectory is determined by a leakage current $I(V_t)$. The capacitor $C_\m{in}$ couples the input signal $x_t$ into the dynamical system. The challenge is that an ultra-low leakage current $I(V_t)$ is required to ensure that the dynamical system is operational for the duration of sensing and data-logging. For instance, a \SI{1}{\V} change across a \SI{1}{\pico\F} on-chip capacitor over a duration of 1 day would require a leakage current of 10 attoamperes. Even if it were possible to implement such low-leakage currents, it is difficult to ensure that the magnitude of the currents match across different devices to ensure state synchronization. The second challenge with regard to data-logging is that there exists a trade-off between the non-linearity in the dynamical systems response and the duration over which the information can be retained. As shown in Supplementary Figure~1a--b, if a constant leakage element (for example reverse leakage current) is used, not only do the system trajectories rapidly converge to the final steady state, but the modulation signal does not cause a change in the sensing system trajectory with respect to the reference system trajectory. On the other hand, a resistive or a direct-tunneling leakage element will be sensitive to the changes in modulation signal but will be unable to keep the two trajectories separated for long periods of time, leading to low retention-time. In this report, we show that a differential G.A.S. dynamical system~\cite{mehta2019differential} implemented using a Fowler-Nordheim (FN) quantum tunneling device~\cite{liangTEDTimer} can address all these challenges.

	\section*{Results}
	\subsection{Differential FN tunneling device acts as a long-term reliable synchronized dynamical system}
	\begin{figure*}
		\centering
		{\includegraphics[width=0.95\columnwidth]{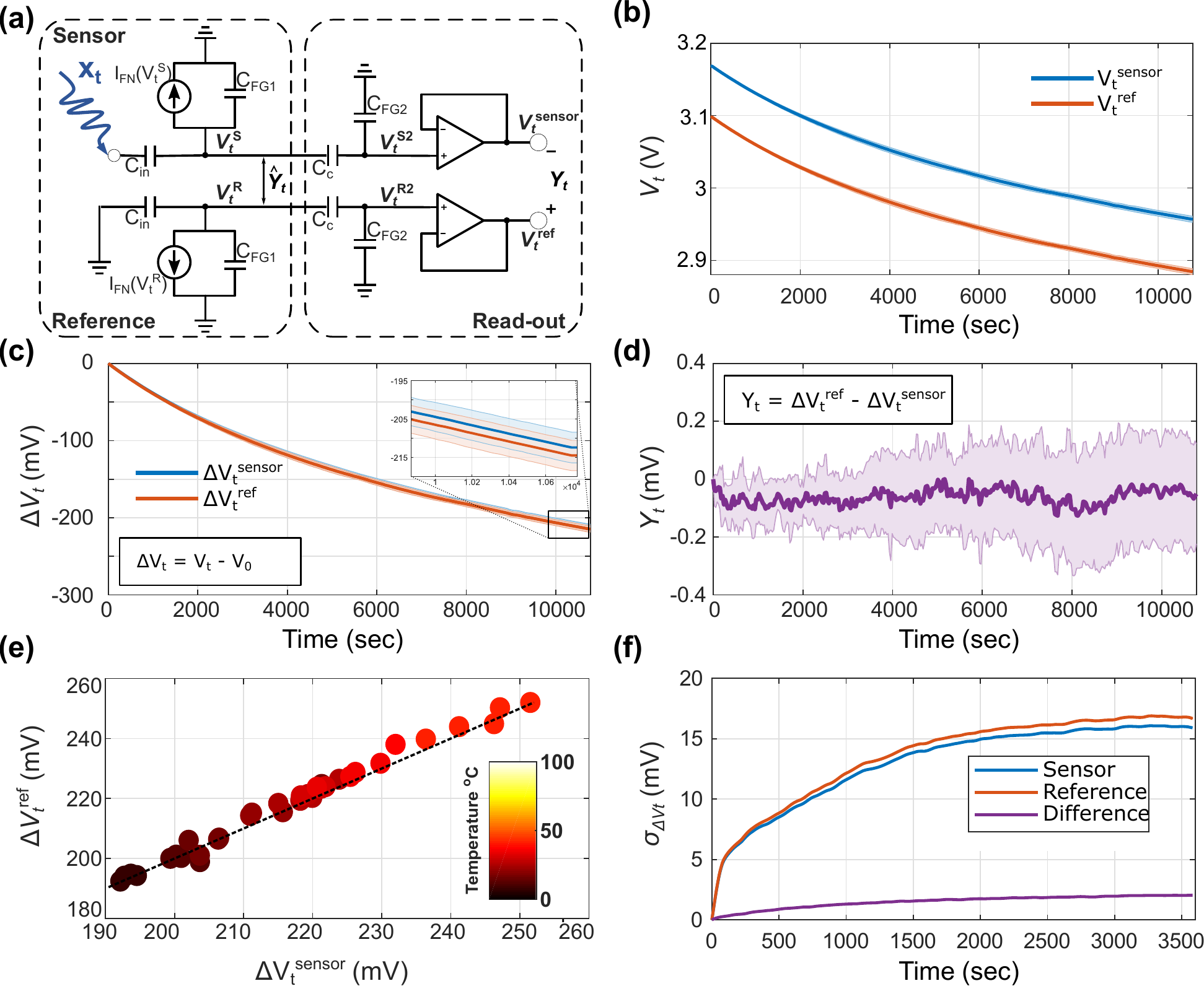}}
		\caption{\textbf{Differential FN sensor-data-logging device and its measured response.} a) Equivalent circuit of the differential FN device coupled to the read-out circuitry. b) Sensor and reference output voltages measured across nine trials after the device is initialized. c) Change in sensor and reference values compared to the initial value $V_0$ as $\Delta V_t = V_t - V_0 $. Shaded region in inset shows $\pm 1$ standard deviation. d) Measured desynchronization between the sensing and reference devices, with bold line showing mean across trials. e) Synchronization measured across a range of operating temperatures (\SI{5}{\celsius} to \SI{40}{\celsius}). The gradient (dark red to yellow) denotes an increase in operating temperature. f) Standard deviation measured for the sensor, reference and the difference over 36 trials and across range of operating temperatures.}
		\label{fig2}
	\end{figure*}
	The operating physics of an FN quantum-tunneling based dynamical system is illustrated using an energy-band diagram in Fig.~\ref{scheme}c~\cite{lenzlinger1969fowler}. From a practical point-of-view, this energy-band configuration can be achieved across a thermally grown gate-oxide (silicon-di-oxide), which acts as an FN tunneling barrier that separates a lightly-doped n-type semiconductor substrate from an electrically insulated but conductive polysilicon island (labeled as a floating-gate). A two-dimensional electron gas and a triangular FN tunneling barrier (as shown in Fig.~\ref{scheme}c) is created by initiating a large potential difference across the semiconductor-floating-gate interface. Thermally-excited electrons then tunnel through the triangular FN tunneling barrier onto the floating-gate ($FG$) and cannot escape due to the surrounding electrical insulation. Each electron that tunnels through the barrier and is retained, changes the potential of the floating-gate which in turn decreases the slope of the FN tunneling barrier (shown in Fig.~\ref{scheme}c). Fig.~\ref{scheme}d shows the cross-section of such an FN tunneling device, whereby the floating-gate is coupled to a programming transistor $P$ and a source follower buffer $B$. The read-out procedure and the procedure to initialize the charge on the floating-gate is described in the Methods section and in Supplementary Figure~2. In~\cite{liangTEDTimer}, we showed that the continuous-time dynamics of this device can be modeled using a first-order differential equation which results in the change in floating-gate voltage $V_t$ at time-instant $t$ as 
	\begin{equation}\label{timerEqn}
		V_t = \frac{k_2}{\log(k_1 t+ k_0)} + k_3
	\end{equation}
	where $k_0$--$k_3$ are model parameters. The parameters $k_1$ and $k_2$ depend on the area of tunneling junction, capacitance, temperature and material properties and the device structure, the parameter $k_3$ depends on the read-out mechanisms and the parameter $k_0$ depends on the initial conditions. For the proposed sensor-data-logger we employ a differential configuration as shown in Fig.~\ref{fig2}a. The initial voltage (equivalently, charge) on each floating-gate is precisely programmed through a combination of tunneling and hot-electron injection (see Calibration and Initialization in Methods)~\cite{harrison2001cmos}. One of the FN device's (labeled as the sensor) dynamics is modulated by an input signal $x_t$, and its desynchronization is measured with respect to a reference FN device as :
	\begin{equation}\label{calcResponse}
	\begin{split}
	\hat{Y}_t = V_t^R - V_t^S
	\end{split}
	\end{equation}
	Here, $V_t^S$ and $V_t^R$ refer to the sensor and reference floating-gate voltages respectively.
	A capacitive divider (formed by $C_\m{c}$ and $C_\m{FG2}$) followed by a source-follower is used to read-out the floating-gate potential through the output node as shown in Fig.~\ref{fig2}a. The floating-node formed at the capacitive divider is independently programmed to a lower value ($\approx$~\SI{3}{V}) to ensure low probability of unwanted tunneling or injection through the transistor $FG_2$.
	The outputs of the sensor and reference nodes, $V_t^\m{sensor}$ and $V_t^\m{ref}$ respectively, are measured using an external data acquisition system (Keithley DAQ6510) and shown in Fig.~\ref{fig2}b. The differential output $Y_t$ in Fig.~\ref{fig2}a is measured with respect to the initial value as
	\begin{equation}\label{calcResponse2}
	Y_t = (V_t^\m{ref} - V_t^\m{sensor}) - (V_0^\m{ref} - V_0^\m{sensor}) = \Delta V_t^\m{ref} - \Delta V_t^\m{sensor} \propto \hat{Y}_t
	\end{equation}
	For calculating $Y_t$, we use the change from their initial voltages at time-instant $t=0$ seconds ($\Delta V_t$ in Fig.~\ref{fig2}c) to eliminate the offset in the read-out stage.
	For each device, less than $1\;\%$ deviation was observed across trials, demonstrating the reliability of the tunneling dynamics and the reliability of the measurement setup. With respect to the differential measurements, $Y_t$ should be \SI{0}{V} in a perfectly synchronized system. However, due to device mismatch and due to differences in the initialization procedure, we observe a baseline drift across all trials. This manifests as variations in device parameters $k_1$--$k_3$, which were estimated by regressing equation~\ref{timerEqn} to the empirical data (Supplementary Table~1). The estimated parameters were then used to compensate for drift and to determine the sensor output (Supplementary Figure~3). Post-drift corrections are shown in Fig.~\ref{fig2}d, which shows the maximum difference between a pair of trials to be less than \SI{300}{\micro\V}.
	We measured the desynchronization of the differential FN device across temperatures ranging from \SI{5}{\celsius} to \SI{40}{\celsius}. Higher temperatures led to faster tunneling, which led to a larger variation in $\Delta V_t$ within the range of \SIrange{200}{260}{\milli\V} as a function of temperature (Fig.~\ref{fig2}e). Despite this variation, the measured desynchronization $Y_t$ had a significantly lower variance with standard deviation below \SI{1}{\milli\V}. These results show that the differential architecture is capable of compensating for variations in temperature. Note that an incorrect initialization of the reference device with respect to the sensor device will make the temperature compensation less robust, as shown by an outlier in Supplementary Figure~4. 
	
	\subsection{A simple behavioral model explains the data-logging principle}
	\begin{figure*}
		\centering
		{\includegraphics[width=0.95\columnwidth]{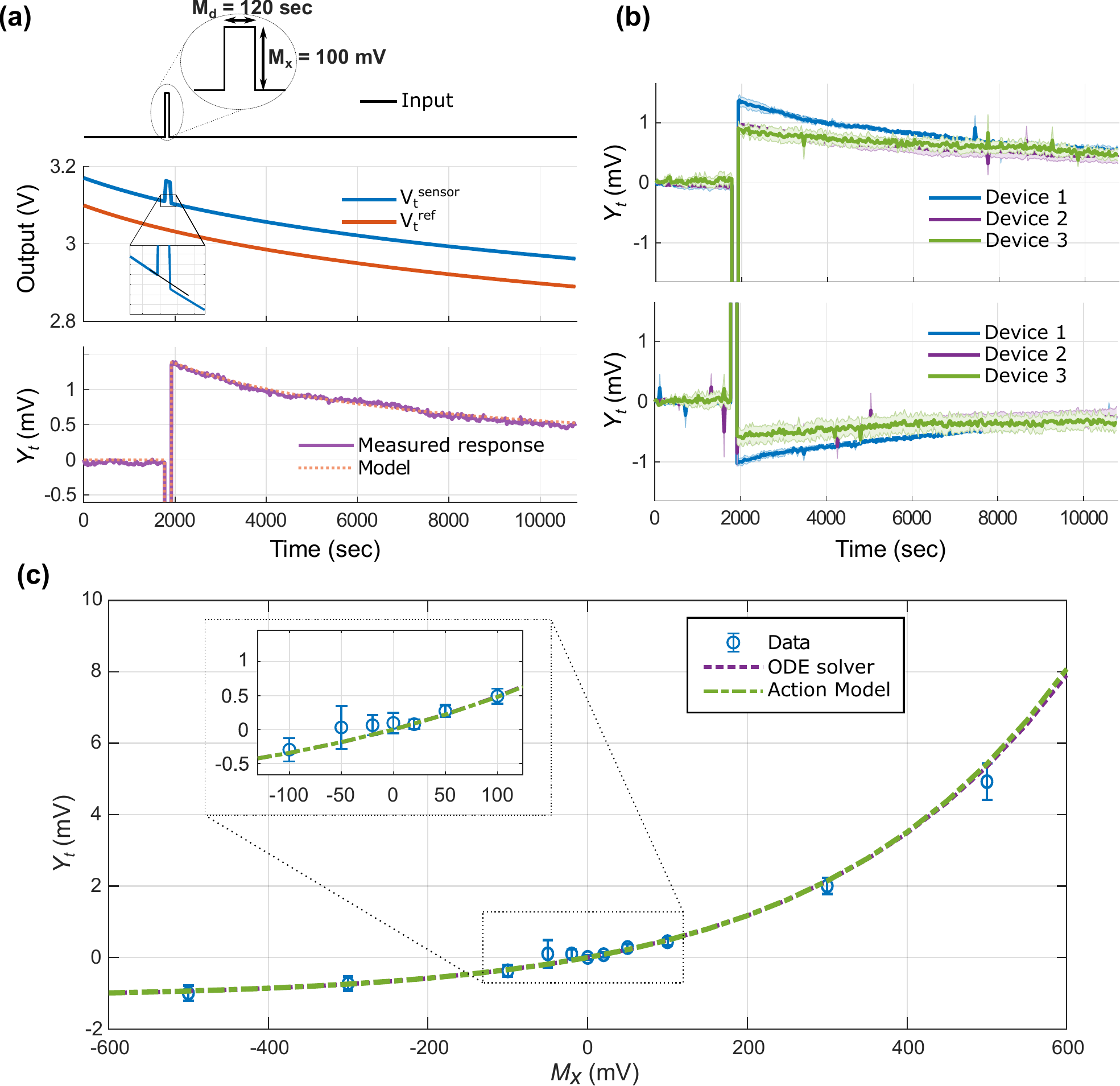}}
		\caption{\textbf{Rectifying response of the sensor-data-logger device.} a) Output measured from the device when subjected to an input pulse. During the positive half of the input pulse, the tunneling-rate increases and desynchronizes the sensor device with respect to the reference device. b) Responses measured from three loggers across three trials. The loggers were initialized to different conditions, hence the difference in their measured responses. c) Sensor responses for input signals over a range of amplitudes. Responses follow an exponential model, which can be accurately modeled by the action model and an ODE solver.}
		\label{recorderExps}
	\end{figure*}

	In the Methods section, we have derived a tractable mathematical model for the data sensed and stored by the sensor-data-logger in response to an arbitrary time-varying input signal $x_t$. We found that the output of the data-logger $Y_T$ measured at time-instant $T$ can be expressed as
	\begin{equation}\label{recorder model}
	Y_T = R(T) A_x(T)
	\end{equation}
	where $A_x(T)$ represents the total ``action'' due to the input signal $x_t$ accumulated up to the time instant $T$ and $R(T)$ is a ``forgetting'' factor that is independent of the input signal $x_t$. $R(T)$ models the data retention capability and arises due to resynchronization of the sensor and reference FN devices, after the sensor device is perturbed by $x_t$. In the Methods section, we show that the action $A_x(T)$ can be expressed in terms of device parameters as
	\begin{equation}\label{Action model}
	A_x(T) =\frac{k_1}{k_2} V_0^2\exp\left(\frac{-k_2}{V_0}\right) \int_{0}^{T} \left[\left(1+\frac{C_R x_t}{V_t}\right)^2 \exp \left(\frac{k_2 C_\m{R} x_t}{V_t(V_t+C_\m{R} x_t)} \right) - 1\right]\diff t
	\end{equation}
	and the resynchronization term $R(T)$ can be expressed as
	\begin{equation}\label{Resync model}
	R(T) = \frac{V_t^2}{V_0^2}\exp\left(\frac{k_2}{V_0} - \frac{k_2}{V_T}\right).
	\end{equation}
	Here $V_t$ is given by equation~\ref{timerEqn} with $V_0$ and $V_T$ representing the device voltage at time-instant $t=0$ and $t=T$ seconds. The parameter $C_\m{R}$ in equation~\ref{Action model} models a capacitive divider that is formed due to the coupling of the input capacitance onto the floating-gate. The Supplementary Figures.~S5 and~S6 show several examples of signals $x_t$ for which the first-order action model given by equation~\ref{recorder model} accurately tracks a more computationally intensive ordinary differential equation (ODE) based device model. In Supplementary Figure~7, we show the ``action'' $A_x(T)$ corresponding to different signal types with different magnitude and energy. The results show that $A_x(T)$ is monotonic with respect to energy and hence can be used as a measure of cumulative energy.
	
	In our controlled experiments we subjected the FN data-logging device to a square pulse of varying magnitude but with a fixed duration of 120 seconds. This duration was chosen because it is sufficiently long enough to elicit a measurable response and for the purpose of device characterization. Also, the pulse was applied at a fixed time (1,800 seconds), after which the desynchronization $Y_T$ was measured at different values of measurement time $T$. Experiments were conducted over a duration of 10,800 seconds (3 hours), with the data-logger responses measured every 30 seconds. Each data-logger was calibrated to similar initial conditions for all experiments wherein the sensor and the reference nodes were initialized to equal tunneling rates. A typical experiment demonstrating the recorder in operation is shown in Fig.~\ref{recorderExps}a, which matches the model described in the Methods section. The RMSE between the model and measured data is \SI{61}{\micro\V} with an $R^2$ of 0.9999.
	
	Measurement results across three repeated trials for input signals of magnitude \SI{100}{\mV} and \SI{-100}{\milli\V} are shown in Fig.~\ref{recorderExps}b. The \SI{100}{\mV} signal resulted in a sensor response of \SIrange[range-phrase=\,--\,]{0.8}{1.5}{\mV} for the three data-logging devices. At the end of three hours, due to resynchronization, the sensor response decreases down to \SIrange[range-phrase=\,--\,]{0.5}{0.6}{\milli\V}. For the \SI{-100}{\mV} input, responses after the modulation were in the range of \SIrange{-0.5}{-0.9}{\mV}, which dropped to \SI{-0.2}{\mV} at the end of three hours. Though the three recorders had different responses, they were consistent across trials for the same recorder. The device responses at the end of three hours for input signals of different magnitudes are shown in Fig.~\ref{recorderExps}c. From the figure, it is evident that the data-logging device response is similar to a rectifier as summarized by the action model in equation~\ref{Action model}. The action model fits the data for this wide range of input conditions with an $R^2$ of 0.9855.
	
	\subsection{Self-powered operation of the proposed device}
\begin{figure*}
		\centering
		{\includegraphics[width=0.95\columnwidth]{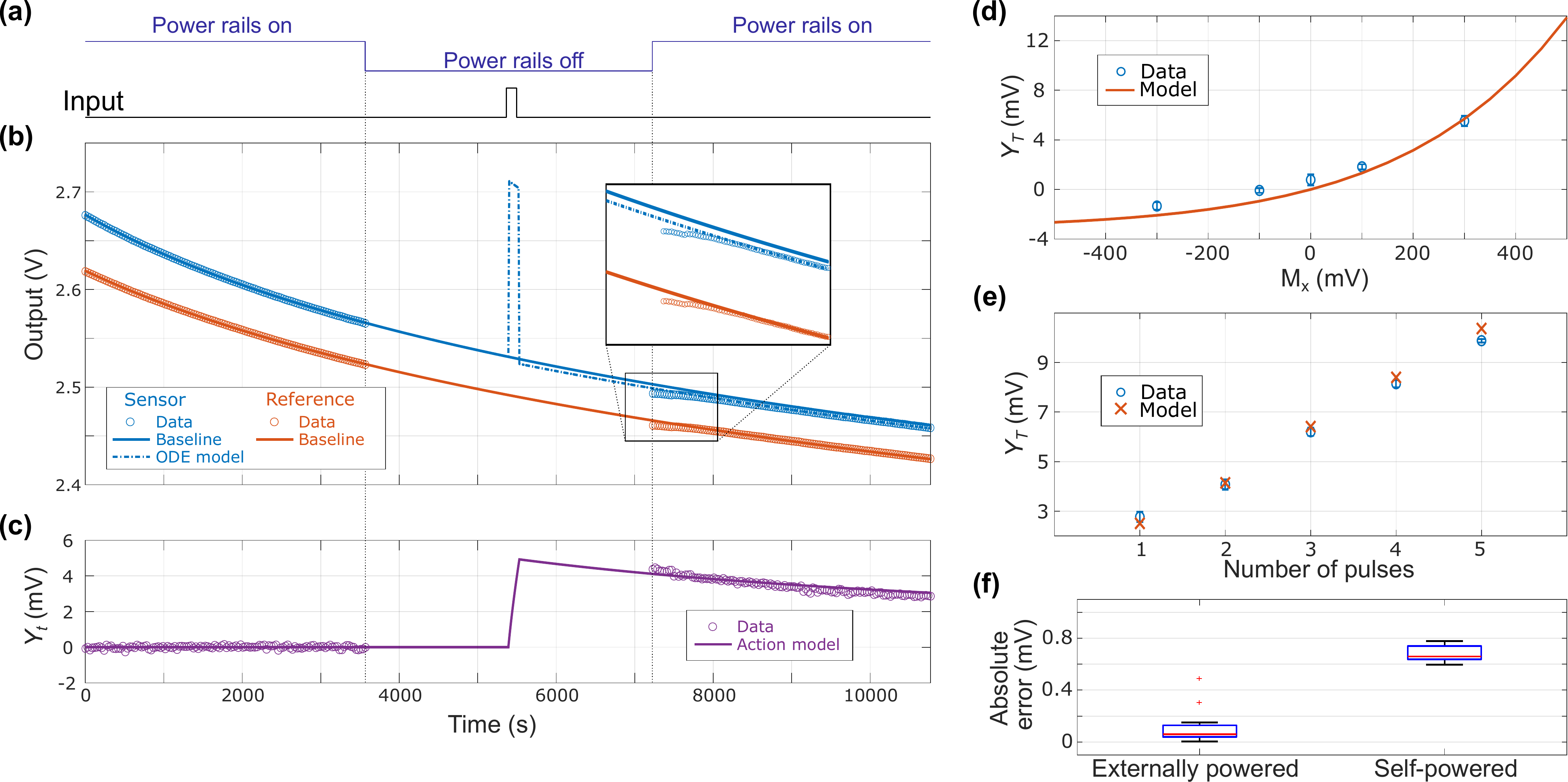}}
		\caption{\textbf{Verification of the proposed device for self-powered sensing and data-logging.} a) Power to the system is switched off at the 1 hour mark, and then turned back on at 2 hour mark. The input pulse is applied at the 1.5 hour mark
			for a duration of \SI{120}{\s} seconds. b-c) Output measured from the recorder when the power is ON and the comparison with the predicted model showing the process of desynchronization. d) Recorder responses for input signals over a range of amplitudes. Responses follow an exponential model, which can be modeled by the action model. e) Recorder responses for varying number of pulses (\SI{400}{\mV} magnitude, \SI{50}{\s} duration each). f) Distribution of absolute errors between measured data and model predictions, for externally powered and self-powered cases estimated across all experiments.}
		\label{selfPoweredExp}
	\end{figure*}

	The self-powered dynamical system created by FN tunneling leakage implies that the device can operate without any external power source, once initialized. We have verified this mode of operation by first disconnecting the sensor-data-logger from any power-supplies and then applying an external signal as an input. The experimental protocol and representative results are shown in Fig. \ref{selfPoweredExp}a. Immediately after powering on the system, the output of the reference node was measured to be lower than the value predicted by the model given by equation~\ref{timerEqn}. However, the measurement stabilized within 200 seconds and the output closely matched the model for the rest of the experiment, indicating that FN tunneling dynamics were conserved in the self-powering mode. Additionally, desynchronization between sensor and reference nodes was observed immediately after power was turned on, indicating that the external signal ``acted'' on the sensing node. Errors introduced during the stabilization period were consistent between the sensor and reference nodes --- the differential architecture attenuated these errors. The magnitude of the response is an exponential function of the input signal magnitude (Fig. \ref{selfPoweredExp}b) as predicted by the action model of equation~\ref{Action model}. Similarly, the recorder was able to record the number (and thus the energy) of discrete pulses applied (Fig. \ref{selfPoweredExp}c) in the power-off state. The mean absolute error between model and observed data was \SI{0.7}{\mV}, higher than the errors obtained for continuously-powered case (Fig. \ref{selfPoweredExp}d).
	
	\subsection{Energy budget, sensing and retention limits}
	The rectification property of the FN data-logging device can be useful for measuring and logging the intensity of a time-varying signal like bio-potentials or accelerometer output. The device is sensitive to input signals of any intensity since there is no threshold requirement on the input signal to activate the sensor. The caveat being, the data retention times for small magnitude signals will be shorter due to the resynchronization (modeled by $R(T)$ in equation~\ref{Resync model}) and operational noise in the recorder. 
	In a perfectly matched differential system, and in the absence of any input, the device response should be exactly \SI{0}{\V}, because of synchronization. However, environmental factors, mismatch between the sensing and reference nodes, or stochasticity in tunneling mechanism, cause desynchronization and the recorder response deviates from the baseline. In general, the variance in the output increases with time (See Fig.~\ref{fig2}d for example). This increase in variance over time is a form of operational noise in the system ($\sigma_t$). A model for $\sigma_t$ could be estimated by letting the recorder operate with respective inputs connected to the ground (similar to input referred noise experiments) and measuring the deviation of output from the baseline. Another source of noise is the readout noise ($N_\m{0}$) which limits the resolution to which charge on the floating gate can be measured. Total noise ($N_t$) is the sum of these two noise sources. While noise increases with time, recorder response decreases due to resynchronization. For a signal of given action, there will be a time instance $T_\m{ret}$, beyond which the signal-to-noise ratio (SNR) goes below a chosen threshold and input signal cannot be reconstructed with a desirable degree of certainty. We chose unity as our threshold for SNR, and we defined data retention as the time at which the signal falls below system noise. The Supplementary Figure~8a shows via an illustration how data retention capacity for a given noise model can be estimated. 
	The Supplementary Figure~8b shows that data retention capacity increases exponentially with the signal action. For a \SI{10}{\mV} action, we could expect to measure significant deviation from baseline for over 300,000 seconds ($\approx$ 4 days). The action model can be used to estimate the energy-budget requirement on the sensing signal. Since the average FN tunneling current is $10^{-17} A$, the energy budget is less than an attojoule. Note that this is the energy to trigger desynchronization. However, isolating the energy dissipated due to FN tunneling from other energy dissipation factors is challenging because the FN tunneling current is on the order of attoamperes, which is orders of magnitude smaller than the reactive current generated by the transducer and the leakage current flowing through ESD protection diodes. In the Supplementary Note~9, we estimate the energy budget when the proposed sensor-data-logger is driven by an arbitrary sensor signal.
	
	Noise in the system can also be described by the effective number of bits (ENOB) (Supplementary Figure~8c). For an assumed action range of \SI{10}{\mV}, 10 bits precision can be initially expected in a system with \SI{10}{\micro\V} readout noise. In a perfectly matched system, ENOB would drop to 0 at $\approx{}2\times{}10^6$ seconds (total recorder lifetime), but with the added operational noise it takes $\approx{}3\times{}10^5$ seconds to reach 0. Readings from multiple recorders can be combined to increase the effective number of bits of the system.
		
	\subsection{Self-powered sensing of action due to ambient acceleration}
\begin{figure*}
		\centering
		{\includegraphics[width=7in]{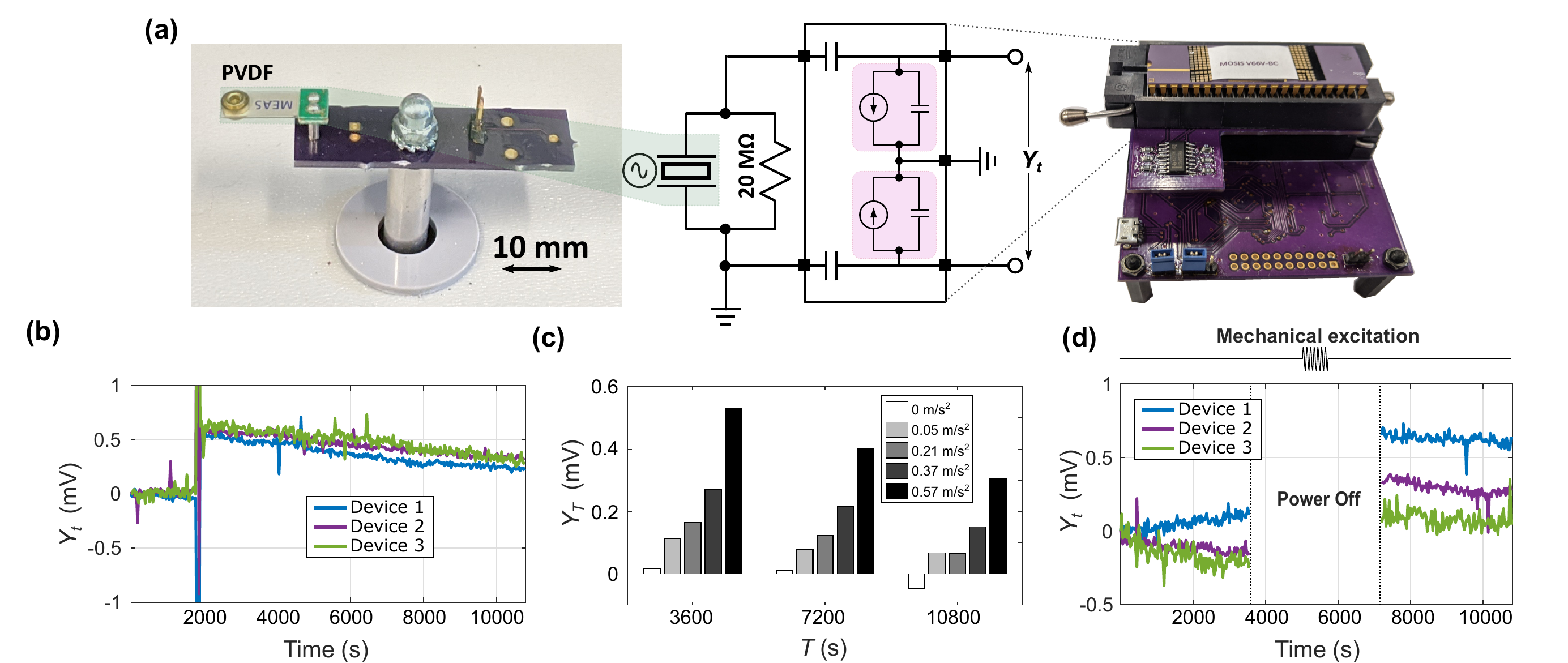}}
		\caption{Self-powered sensing and data-logging of mechanical acceleration. a) Experimental setup showing a piezoelectric (PVDF) transducer connected to the FN sensor-data-logger chipset. b) Logger response when \SI{58.6}{\milli\g} (\SI{0.57}{\m\per{\sec\squared}}) acceleration was applied to the piezo cantilever (gain of \SI{6}{\V\per\g} at \SI{75}{\Hz} resonant frequency) at \SI{72}{\Hz} for 100 sec. c) Recorder responses at different readout times for a range of input frequencies. All modulated responses were statistically different from the unmodulated case at all readout times. d) Recorder was powered off in the shaded region. During that time period, vibration table was actuated, which was recorded as evidenced by the recorder value when power supply was turned on.}
		\label{figPiezo}
	\end{figure*}

	In this section, we demonstrate the use of the proposed sensor-data-logger for battery-free sensing of ambient acceleration. We chose a piezoelectric transducer for sensing mechanical acceleration and for directly powering the sensor-data-logger device. Note that in this regard, other transducers for e.g. photodiodes, RF antennas, thermocouples could also be directly interfaced to the FN data-logging device to create other self-powered sensing modalities. A schematic of the experimental setup is shown in Fig.~\ref{figPiezo}a. A PVDF (polyvinylidene difluoride) cantilever [TE Connectivity's Measurement Specialties MiniSense 100 Vibration Sensor with nominal resonant frequency – \SI{75}{\Hz}] was mounted on a benchtop vibration table [3B Scientific Vibration Generator - U56001] that is externally actuated by a function generator. The table was actuated at an off-resonant frequency of \SI{72}{\Hz} for a range of actuating amplitudes. We simultaneously measured acceleration using a 3-axis accelerometer [Adafruit LIS3DH accelerometer] to use as the ground truth. Results are shown in Fig.~\ref{figPiezo}b--c. We observed significant responses for vibration signals down to an acceleration of \SI{0.0052}{\g} (\SI{0.05}{\m\per{\sec\squared}}). For context, a refrigerator vibrates with an acceleration of around \SI{0.1}{\m\per{\sec\squared}}~\cite{roundy2005effectiveness}. The expected maximum output power of the piezoelectric sensor is on the order of tens of nanowatts of which only a fraction is used by the recorder to store the information. In the final experiment, we electrically disconnected all power to the recording system at the 1 hour (3,600s) mark, actuated the vibration table at 1.5 hours (5,400 s) and reconnected the system at 2 hours (7,200 s) to readout the output of the data-logger. We observed vibration-induced desynchronization in this set of experiments as well, with the deviation as expected based on the earlier characterization tests.
	
	\section*{Discussion}
	In this report, we proposed a novel method for designing an ultra-energy-efficient sensor-data-logging device, where the energy of the sensing signal is used to modulate the state trajectories of a synchronized dynamical system. We showed that a Fowler-Nordheim (FN) quantum tunneling device~\cite{liangTEDTimer} can be used to implement the proposed sensor-data-logger on a standard silicon process. 

	Our modeling study summarized in Supplementary Figure~9 shows that there are multiple parameters, both operational and design parameters, that affect the retention time (or resynchronization) of the FN device. Change in any parameter that increases (decreases) the ``action'' of the signal, would also lead to faster (slower) resynchronization. Thus, its net effect on the system depends on the total duration for which the input signal was applied. The initial charge on the floating-gate and time to sample are operational parameters as they can be set at run-time, as required by the specific application. Larger time intervals allow the input signal to be integrated over a longer period of time, but it does not change the sensitivity to the signal. For a signal of given action, the measured value decreases as $T$ increases due to resynchronization (Fig.~9a). Initializing the device to a higher voltage leads to higher sensitivity but only up to a certain limit (Fig.~9b). The reason is that higher sensitivity also leads to faster resynchronization determined by design parameters $k_1$ and $k_2$.
	$k_1$ can be tuned by varying the area of tunneling junction and capacitance sizing. We found that increasing the area or lowering the capacitance would increase the sensitivity of the system but only to a certain point (Fig.~9c). Beyond this point, the gains are only marginal, at the expense of a larger footprint. Moreover, the capacitance is a function of the tunneling junction area and thus the ratio of area to capacitance is bounded and depends on the permittivity of the insulator. Smaller oxide thickness would decrease $k_2$ and sharply increase the sensitivity (Fig.~9d). However, at these scales the effect of other processes like direct tunneling cannot be ignored. Exploring different materials could have significant impact on both $k_1$ and $k_2$, as they affect the parameters $\alpha$ and $\beta$ (equations~\ref{currentDensity1} and~\ref{params1}). When the input signal is a single pulse, the time-of-occurrence of the pulse also plays a role in the measured response, as shown in Supplementary Figure~11. However, this effect is weaker than that of other factors.
	
	The desynchronization based approach reduces the energy budget required for data-logging, we estimate that the proposed device can operate at an energy budget lower than an attojoule while retaining the information for at least 3 hours. In standard analog sensor circuits, quiescent current is sourced from a power source for continuous operation. 
	In the proposed device, the quiescent current is the FN quantum tunneling current, which is sourced from the pre-charged capacitor and ambient thermodynamic fluctuations. Hence, no external power source is required for operation. For modulating the sensor, energy is extracted from the signal being sensed. If the energy dissipated at the input signal source (due to finite source impedance) is ignored, the energy budget required to modulate the state of the FN device is less than \SI{100}{\atto\J}. In practice, the energy  from the source is spent on charging the capacitor, and for maintaining DC voltage at the source, as described in Supplementary Note~9. For example, when the magnitude of the input signal is \SI{100}{\mV}, \SI{15}{\femto\J} is used for charging a \SI{300}{\femto\F} input capacitor. The DC input impedance of the proposed device was measured to be greater than \SI{{}e17}{\ohm}; thus, the energy required to maintain a voltage potential of \SI{100}{\mV} for 120 sec is less than \SI{100}{\atto\J}. Many signals of interest have power levels greater than this, and can provide sufficient energy for modulating the sensor, provided the system impedance is matched to the source.
	
	However, for time-varying sensor signals, a more efficient power transfer or sensing is achieved as some energy stored in reactive elements like capacitors can be recovered. In Supplementary Figure~10, an equivalent circuit model corresponding to the FN tunneling device is shown along with a simplified sensor-transducer equivalent model. We note that the input impedance of the recorder is predominantly capacitive and the only dissipative factor arising during sensing/data-logging is due to the FN tunneling current. The equivalent circuit model allows estimation of the power dissipated by a device that is excited by an arbitrary sensor signal. In Supplementary Note~9, we show that a broadband AC input signal with upper cutoff frequency of \SI{1}{\kilo\Hz} and amplitude of \SI{100}{\mV} has an estimated energy dissipation by the system of \SI{5}{\atto\J} for an event lasting 100 seconds.

	Using FN quantum tunneling to implement the dynamical system has some key advantages. Its stability allowed us to create a pair of synchronized devices which is compensated for environmental variations. Its predictability was used for modeling, and we were able to derive a recorder response model that matched experimental data with $98.8\;\%$ accuracy. Its dynamics follow a $1/\log(t)$ characteristic, which yields a long operational life. The non-linear response leads to rectification of input signals and offers an opportunity for time stamping and reconstruction. A more rigorous and theoretic investigation into the use of dynamical systems for information reconstruction will be the topic of future research.
	
	At its core, the proposed device consists of four capacitors and two transistors (4C-2T), and can be implemented on any standard CMOS process. The current design is a proof-of-concept and is not optimized for sensitivity or form factor. Modeling analysis in Supplementary Figure~9 shows that both of these parameters can be improved by minimizing the capacitance, while maintaining the capacitance ratio ($C_\m{R}$). To achieve this, an optimum balance between the input capacitor, decoupling capacitor and parasitic capacitance at the poly-substrate tunneling junction needs to be obtained. Better matching of the sensor and reference nodes (tunneling junctions, capacitors and readout circuits) using advanced analog layout techniques should be able to reduce the operational noise in the recorder and thereby increase the data retention capacity. Readout and common-mode noise can be further reduced by implementing a low-noise on-chip instrumentation amplifier. Multiple units of independent recorders could be used to increase the SNR of the recordings.

	Any passive sensor that is capable of transducing a physical signal into an electrical signal (voltage or current) can be interfaced with our system. These include piezoelectric transducers, photodiodes, radio-frequency antennas, thermocouples, triboelectric generators etc. Passive sensors like strain gauges, that do not directly produce electrical output but instead effect a change in resistance are not compatible with our system. Similarly, many chemical transducers like dopamine sensors that require an activation voltage (external biasing or power) are also not applicable for self-powered data-logging. However, chemical sensors like amperometric glucose sensor that have the ability to generate electrical charge during the process of sensing should work with our system. In addition, there can be practical issues in measuring certain types of signals. For example, the limited action generated by signals like neural action-potentials may not be measurable due to resynchronization and system-noise.
	Finally, the proposed recorder could be directly integrated with FET (field-effect transistor) based sensors~\cite{siSensor_middelhoek2000celebration,isfet_bergveld2003thirty,ofet_torsi2013organic}, which have been developed for a wide range of applications. As there are no extrinsic powering requirements, there is the potential of integrating these devices on ``smart dust'' platforms as well~\cite{smart_dust_warneke2001smart,neuralDust_seo2016wireless}.
		
	In conclusion, we have described a self-powered sensor-data-logger device that records a cumulative measure of the sensor signal intensity over its entire duration. To achieve this, we designed a pair of synchronized dynamical systems whose trajectories are modulated by an external signal. The modulation leaves its trace by desynchronizing one of the synchronized pairs. The total cumulative measure or action is stored as a dynamical state which is then measured at a later instant of time. The self-powered dynamical system was designed by exploiting the physics of Fowler-Nordheim quantum tunneling in floating gate transistors. We modeled the response of our system to an arbitrary signal and verified the model experimentally. We also demonstrated the self-powered sensing capabilities of our device by logging mechanical vibration signals produced by a small piezoelectric transducer, while being disconnected from any external power source.

	\section*{Methods}
	\subsection*{One-time programming}\label{Methods_prog}
	For each node of each recorder, the readout voltage was programmed to around \SI{3}{\V} while the tunneling node was operating in the tunneling regime. This was achieved through a combination of tunneling and injection. Specifically, VDD was set to \SI{7}{\V}, input to \SI{5}{\V} and the program tunneling pin was gradually increased to \SI{23}{\V}. Around \SIrange{12}{13}{\V}, the tunneling node’s potential would start increasing. The coupled readout node’s potential would also increase. When the readout potential went over \SI{4.5}{\V}, electrons would start injecting into the readout floating gate, thus ensuring its potential was clamped below \SI{5}{\V}. After this initial programming, VDD was set to \SI{5}{\V} for the rest of the experiments.

	\subsection*{Calibration}\label{Methods_calibration}
	After one-time programming, input was set to \SI{0}{\V}, Vprog to \SI{21.5}{\V} for 1 minute and then the floating gate was allowed to discharge naturally. Readout voltages for the sensor and reference nodes were measured every 30 seconds, for 3 hours. The rate of discharge for each node was calculated; and a state where the tunneling rates would be equal was chosen as the initial synchronization point for the remainder of the experiments.

	\subsection*{Initialization}\label{Methods_init}
	Before the start of each experiment, floating gates were initialized to the initial synchronization point, estimated in the previous section. This was done by either setting the input to stable DC point through a digital to analog converter (DAC) or if the DAC value needed was beyond its output limit, then the potential would be increased by setting Vprog pin to \SI{21}{\V}.

	\subsection*{Model derivation}\label{Methods_Tmodel}
	FN tunneling current density $J_\m{FN}$ across a triangular barrier can be expressed as a function of the electric field $E$ across the barrier~\cite{lenzlinger1969fowler}:
	\begin{equation}\label{currentDensity1}
	J_\m{FN}(E)=\alpha {E}^2 \exp(-\beta /E)
	\end{equation}
	where $\alpha$ and $\beta$ are process and device specific parameters~\cite{lenzlinger1969fowler}.
	
	Thus, for a tunneling junction with cross-sectional area $A$ and thickness $t_\m{ox}$, the tunneling current $I_\m{FN}$ for a time-varying
	voltage $V_t$ is given by 
	\begin{equation}\label{currentEqn}
	I_\m{FN}(V_t)= A\alpha (V_t/t_\m{ox})^2 \exp(-\beta t_\m{ox}/V_t).
	\end{equation}
	
	Referring to the equivalent circuit in Fig.~\ref{fig2}a, the dynamical system model when the sensing signal $x_t$ is absent is given by
	\begin{equation}\label{dischargeEqn}
	I_\m{FN}(V_t)= - C_\m{total} \frac{\diff V_t}{\diff t}
	\end{equation}
	where $C_\m{total} = C + C_\m{in}$ is the total capacitance at the floating-gate node.	
	The solution of the equation can be expressed as :
	\begin{equation}\label{timerEqn1}
	V_t = \frac{k_2}{\log(k_1 t+ k_0)}
	\end{equation}
	where
	\begin{equation}\label{params1}
	k_1 = \frac{A \alpha \beta}{Ct_\m{ox}} \: \: \: k_2 = \beta t_\m{ox}
	\end{equation}
	depend on material properties and device structure, while
	
	\begin{equation*}\label{params21}
	k_0 = \exp \left(\frac{k_2}{V_0}\right)
	\end{equation*}
	depends on the initial conditions.
	
	Now, let 
	\begin{equation}\label{dVdt}
	\m{f}(V_t)= -\frac{I(V_t)}{C_\m{total}} = -\frac{k_1}{k_2} V_t^2 \exp\left(\frac{-k_2}{V_t}\right)
	\end{equation}
	
	Desynchronization between the sensor and reference nodes shown in Fig.~\ref{fig2}a occurs because of differences in rates of tunneling, which are caused by differences in electric potentials across the respective floating-gates
	\begin{equation}\label{diffEqn1}
	\begin{split}
	\frac{\diff Y_t}{\diff t} & =  \frac{I_\m{FN}(V_t^S)}{C_\m{total}} - \frac{I_\m{FN}(V_t^R)}{C_\m{total}} \\
	& = \m{f}(V_t^R) -  \m{f}(V_t^S)
	\end{split}	 
	\end{equation}
	
	The reference node $V_t^R$ follows the dynamics of equation~\ref{timerEqn1} as it is not under the action of an external field. Thus, $V_t^R = V_t$. The potential across the sensing node is given by how much it has desynchronized from the reference node ($V_t^R - Y_t$) and the effect of the external field, $x_t$, through the input capacitor $C_\m{in}$.
	
	\begin{equation}\label{VS}
	V_t^S = V_t + C_\m{R}x_t - Y_t
	\end{equation}
	
	where $C_\m{R}$ is the coupling ratio due to capacitive divider formed by $C_\m{in}$ and $C_\m{fg}$.
	\begin{equation}\label{CR}
	C_\m{R} = \frac{C_\m{in}}{C_\m{total}} ; C_\m{total} = C_\m{in} + C_\m{FG1} + C_\m{C}||C_\m{FG2}
	\end{equation}
	
	Substituting $V_t^R$ and $V_t^R$ in equation~\ref{diffEqn1}
	\begin{equation}\label{ode}
	\frac{\diff Y_t}{\diff t} =  \m{f}(V_t) - \m{f}(V_t + C_\m{R}x_t - Y_t)
	\end{equation}
	The above equation is the constitutive differential equation and can be solved using numerical methods for any input signal. To obtain an explicit expression for estimating the response $Y_t$, we assume that $Y_t{\ll}V_t$ and $\textrm{E}(x_t) = 0$ for all $t$, and use Taylor series expansion with first order approximation.
	\begin{equation}\label{odeB}
	\begin{split}
	\frac{\diff Y_t}{\diff t} & = \m{f}(V_t) - \m{f}(V_t + C_\m{R}x_t) + \frac{\diff (\m{f}(V_t))}{\diff V_t}Y_t \\
	\frac{\diff Y_t}{\diff t} & - \frac{\diff (\m{f}(V_t))}{\diff V_t}Y_t = \m{f}(V_t) - \m{f}(V_t + C_\m{R}x_t) 
	\end{split}
	\end{equation}
	
	Multiplying both sides of equation~\ref{odeB} by $1/\m{f}(V_t)$, substituting $\diff V_t{=}\m{f}(V_t) \diff t $ (from equations~\ref{dischargeEqn} and~\ref{dVdt}) and simplifying:
	\begin{equation}\label{ode3}
	\begin{split}
	\frac{\diff Y_t}{\m{f}(V_t) \diff t} - \frac{\diff (\m{f}(V_t))}{{\m{f}(V_t)}^2 \diff t}Y_t & = \frac{1}{\m{f}(V_t)}(\m{f}(V_t) - \m{f}(V_t + C_\m{R}x_t)) \\
	\frac{\diff }{\diff t} \left(\frac{Y_t}{\m{f}(V_t)}\right) & = 1- \frac{\m{f}(V_t + C_\m{R}x_t)}{\m{f}(V_t)}
	\end{split}	 
	\end{equation}
	Integrating both sides with respect to $\diff t$ between the limits 0 and $T$:
	\begin{equation}\label{odeSol}
	\begin{split}
	\frac{Y_T}{\m{f}(V_T)} - \frac{Y_\m{0}}{\m{f}(V_\m{0})} & = \int_0^T{\left(1- \frac{\m{f}(V_t + C_\m{R}x_t)}{\m{f}(V_t)}\right)} \diff t \\
	\frac{Y_T}{\m{f}(V_T)} & = \int_0^T{\left(1- \frac{\m{f}(V_t + C_\m{R}x_t)}{\m{f}(V_t)}\right)} \diff t \\
	Y_T & = \m{f}(V_T) \int_0^T{\left(1- \frac{\m{f}(V_t + C_\m{R}x_t)}{\m{f}(V_t)}\right)} \diff t 
	\end{split}	 
	\end{equation}
	
	Substituting $\m{f}(V_t)$ from equation~\ref{dVdt} into equation~\ref{odeSol}
	\begin{equation}\label{odeSolFN}
	\begin{split}
	Y_T & = \frac{k_1}{k_2} V_T^2 \exp\left(\frac{-k_2}{V_T}\right) \int_{0}^{T} \left[\left(1+\frac{C_\m{R} x_t}{V_t}\right)^2 \exp \left(\frac{k_2 C_\m{R} x_t}{V_t(V_t+C_\m{R} x_t)} \right) - 1\right] \diff t
	\end{split}	 
	\end{equation}
	
	\subsection*{Data availability}
	All the software and experimental data used for generating the figures have been deposited in a public repository (\url{https://doi.org/10.6084/m9.figshare.12814592.v1})~\cite{mehta_chakrabartty_2020}.

	\section*{Acknowledgements}
	This work was supported in part by NIH research grants 1R21EY028362-01 and 1R21AR075242-01. The authors acknowledge the help and resources provided by Prof. Srikanth Singamaneni and Prashant Gupta in acquiring micrographs of the fabricated chips. We thank Dr. Liang Zhou for useful discussions regarding quantum tunneling dynamics and circuit design. Owen Pochettino is acknowledged for helping build a chip testing station.
	
	\section*{Author contributions}
	D.M., K.A. and S.C. conceived the project. D.M., and S.C. designed the experiments. D.M. and K.A. developed instrumentation for data collection. D.M. performed the experiments and collected data. D.M. and K.A. analyzed the data and generated figures. S.C. supervised all aspects of the work. All authors contributed to the writing of the manuscript.
	
	\section*{Competing interests}
	The authors declare no competing interests.
	
	\clearpage
	\setcounter{equation}{0}
	\renewcommand{\figurename}{Supplementary Figure}
	\setcounter{figure}{0}
	\section*{Supplementary Information}
	\subsection{Different leakage mechanisms for dynamical logging devices}
\begin{figure}[h]
		\centering
		{\includegraphics[width=6.5in]{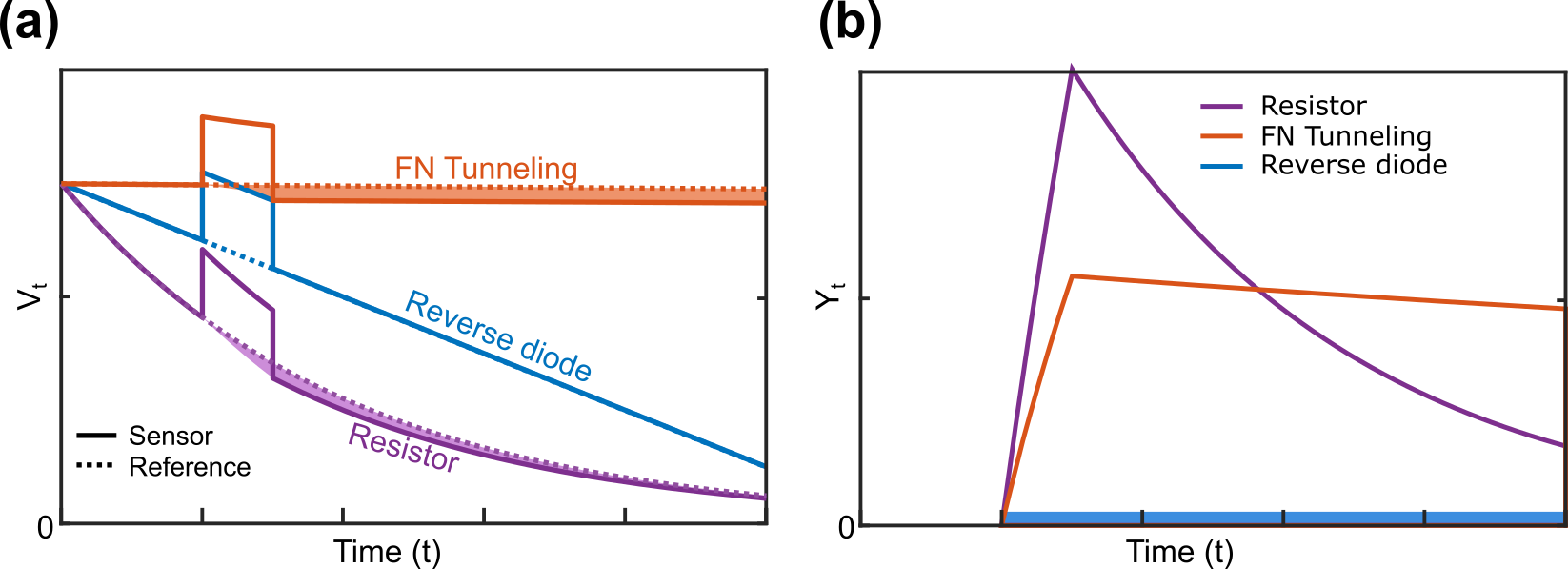}
		}
		\caption{a) Different leakage elements (Fowler-Nordheim tunneling, reverse biased diode and resistor) elicit different resynchronization responses. b) Note that the desynchronization response for the reverse diode case is zero.}
		\label{SI_curves}
	\end{figure}

	Three different types of dynamical systems are simulated based on different leakage element $I(V_\m{t})$ in Fig. 1b. All systems would respond to an external signal (a square pulse), and then resynchronize to their baseline response. This is illustrated in Fig.~\ref{SI_curves}a for three different leakage elements. When the leakage elements is a resistor, the dynamics follow an exponential characteristic. However, an extremely large resistance would be required to sustain the effects of the input pulse (or transient response). As an example, for a system with $C=$ \SI{1}{\pico\F}, $R=$ \SI{1}{\tera\ohm}, $V_0=$ \SI{3}{\V}, a 1-second-long, \SI{100}{\mV} input signal will elicit a response that can be observed for 5.5 seconds. This is illustrated in Fig.~\ref{SI_curves}b. Note that when the leakage element is a constant current (reverse biased diode leakage), the input pulse does not elicit any change in the response. For the leakage element based on FN tunneling, which follows a $1/\log(t)$ dynamics, the input pulse elicits a response that shows a much longer resynchronization time, as shown in Figs.~\ref{SI_curves}a--b. This feature has been modeled and experimentally verified in the main text.
	
	\clearpage
	\subsection{Programming and synchronization}\label{SI_Prog}
	The differential sensor-data-logging system consists of two nodes: sensor and reference node. Each node contains two floating gates decoupled via a capacitor. The charge on the four gates of the system can be individually programmed using a combination of tunneling (increases charge, course) and hot electron injection (decreases charge, fine). The programming block for each gate is selected via a switch. Injection is initiated by setting $V_\m{DD}$ = \SI{7}{\V}, and setting the input pin to a value (via a DAC), such that $V_\m{DS}$ is above \SI{4.2}{\V}. $V_\m{DS}$ can be modulated via the gate voltage because the PMOS is in a source follower configuration. Tunneling is realized by bring $V_\m{tun}$ to a high potential. For programming the tunneling node to be in the FN tunneling regime, we used $V_\m{tun} =$ \SI{21}{\V}. Except for the self-powered experiments using piezo crystals, we did not have to program the tunneling node in FN tunneling regime using $V_\m{prog}$  pin. Instead, we could set the input pin to a stable voltage (analog ground) which would push the node into FN regime. The DAC voltage was calculated for each run such that a tunneling node's potential at the start of each experiment was the same (as measured by the readout node). When the needed DAC voltage exceeded \SI{5}{\V}, we would initiate tunneling. This process allowed us to limit the number of high voltage tunneling cycles and increase the experimental life of the recorder. This process cannot be done in actual deployment because there would not be an external DC source. Hence, for self-powered piezo experiments, we carried out tunneling for each trial. 
	\begin{figure}[ht]
		\centering
		{\includegraphics[width=6.5in]{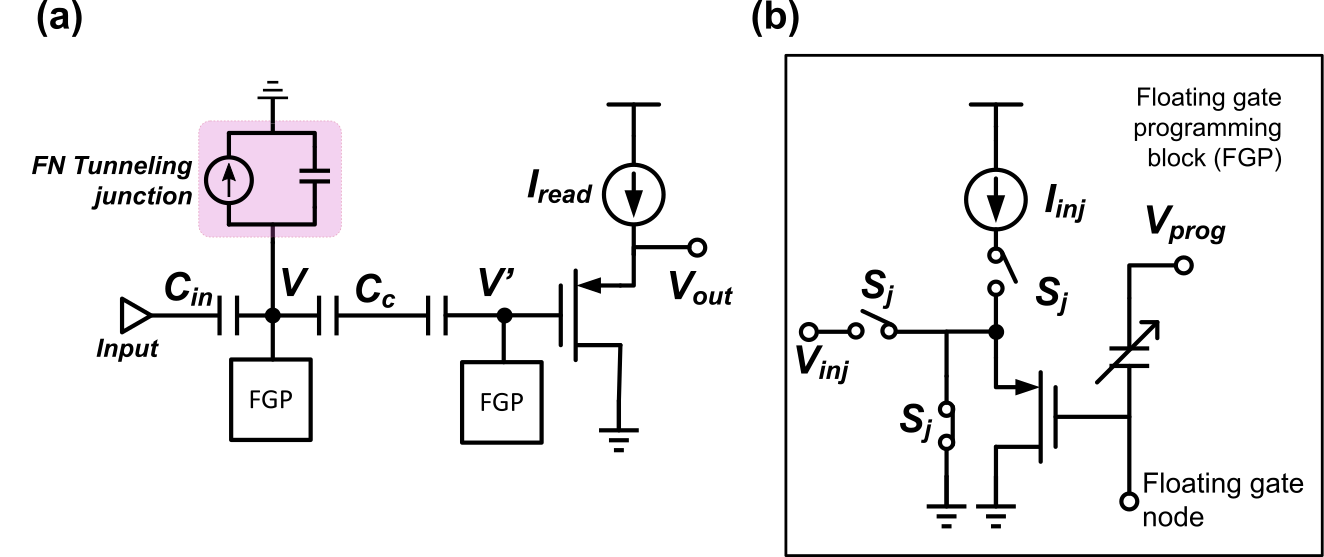}}
		\caption{Programming and synchronization: a) Each floating gate node can be individually programmed through the Floating Gate Programming (FGP) block. b) Electrons can be tunneled out of the floating gate by setting the $V_\m{prog}$ to a high potential. Electrons can be injected into the floating gate via hot electron injection. The switch $S_j$, set via a shift register, allows for individual control of the injection channel. $V_\m{inj}$ node is monitored during injection.}
		\label{SI_circuit}
	\end{figure}	
	
	\clearpage
	\subsection{Device parameters and drift correction}
	Device parameters from equation~1 (Main text) can be experimentally obtained by allowing the floating gate to discharge via Fowler-Nordheim tunneling and fitting the model on observed data.
	
	We obtained the following parameters:
	\begin{table}[h]
		\centering
		\def\arraystretch{1.2}
		\begin{tabular}{r| c|c|c|c } 
			Device No. & Node & $\log (k_1)$ & $k_2$ & $k_3$ \\ 
			\hline \hline
			\multirow{2}{1cm}{1} & Sensor & 38.59 & 347.20 & -4.24 \\ 
			& Ref& 39.47 & 359.04 & -4.43 \\ 
			\hline
			\multirow{2}{1cm}{2} & Sensor & 44.53 & 425.01 & -4.86 \\ 
			& Ref & 42.06 & 389.18 & -4.57 \\ 
			\hline
			\multirow{2}{1cm}{3} & Sensor & 42.14 & 381.26 & -4.35 \\ 
			& Ref & 41.16 & 370.20 & -4.30 \\ 
		\end{tabular}
		\vspace{0.5ex}
		\caption{Device Parameters}
		\label{table:1}
	\end{table}
	
\begin{figure}[h]
		\centering
		{\includegraphics[width=6.5in]{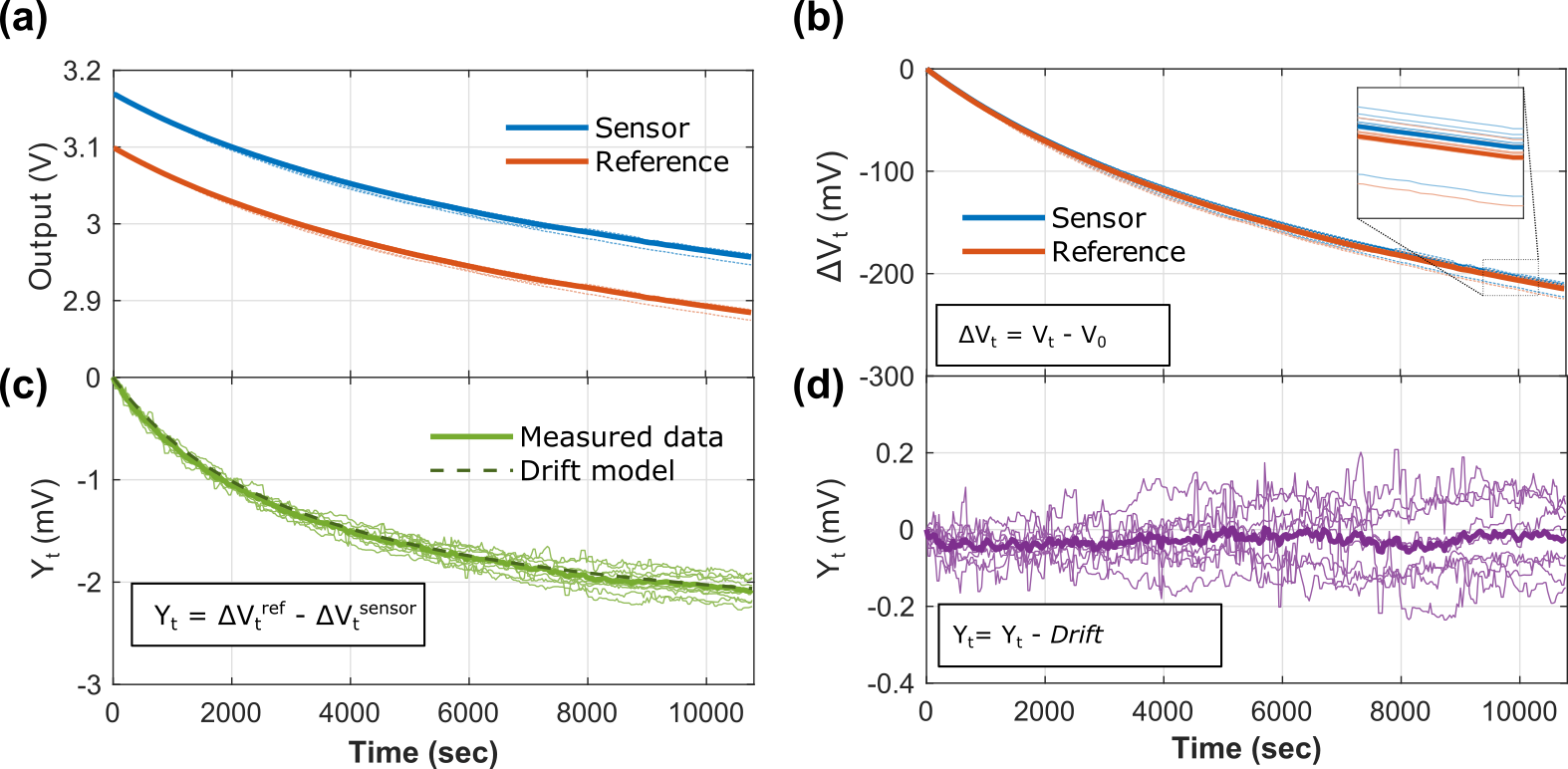}
		}
		\caption{Synchronization after correcting for drift: a) Experimentally measured values of sensor and reference output voltages. b) Change in sensor and reference values from the baseline $\Delta V_\m{t} = V_\m{t} - V_0 $. c) Desynchronization between the sensing and reference nodes. A consistent drift is observed across trials which can be compensated. d) Desynchronization compensated for drift, which is the final recorder response.}
		\label{SI_Sync}
	\end{figure}

	$k_0$ depends on the initial conditions. The starting voltages for each node can be chosen such that the sensor and reference have the same rates and are thus synchronized.  However, the mismatch in other parameters causes the two nodes to drift. The drift is predictable and can be corrected as shown in Fig.~\ref{SI_Sync}. $k_3$ was subsumed into $V_\m{t}$ by setting $V_\m{t} \to V_\m{t} - k_3$. The modified $V_\m{t}$ is used for derivation of the explicit model in equation~4.

	\clearpage
	\subsection{Temperature compensation}
\begin{figure}[h]
		\centering
		{\includegraphics[width=6.5in]{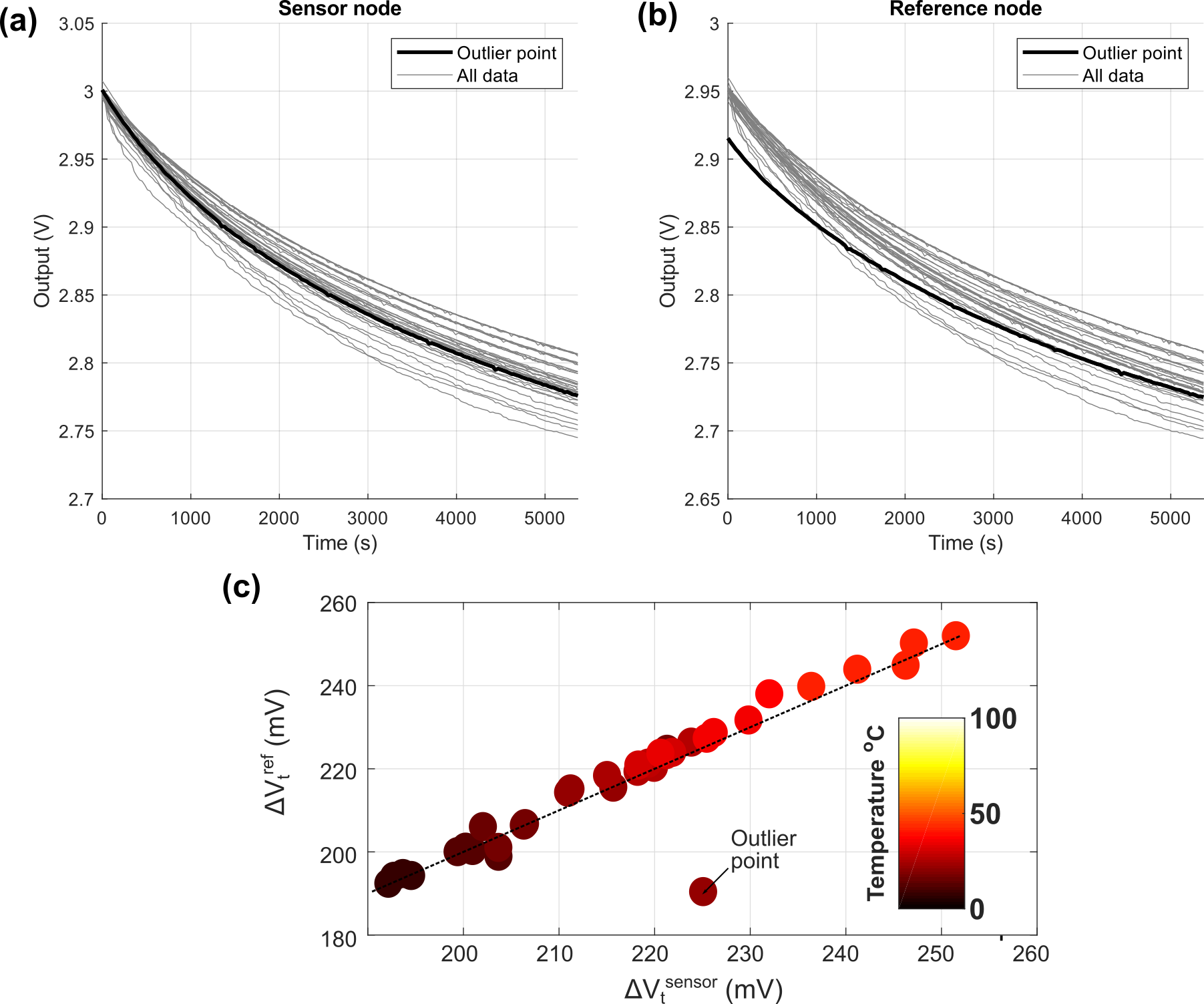}
		}
		\caption{a-b) Raw sensor and reference data for temperature compensation experiments highlighting the outlier which results in incorrect temperature compensation, as shown in c)}
		\label{tempOutlier}
	\end{figure}

	To ensure accurate temperature compensation, it is important to accurately initialize the reference and the sensor devices. To illustrate this we show raw measured data from our experiments (Fig.2) for the sensor and the reference devices. The dataset shows an outlier due to incorrect initialization of the reference device (as highlighted in Fig.~\ref{tempOutlier}, where the reference node was incorrectly initialized lower than the target \SI{50}{\mV} difference from the sensor node). This difference results in improper compensation of the temperature variations. To ensure consistency, we have removed the outlier from Fig.2e. 
	
	\clearpage
	\subsection{Model validation}\label{SI_model}
\begin{figure}[h]
		\centering
		{\includegraphics[width=6in]{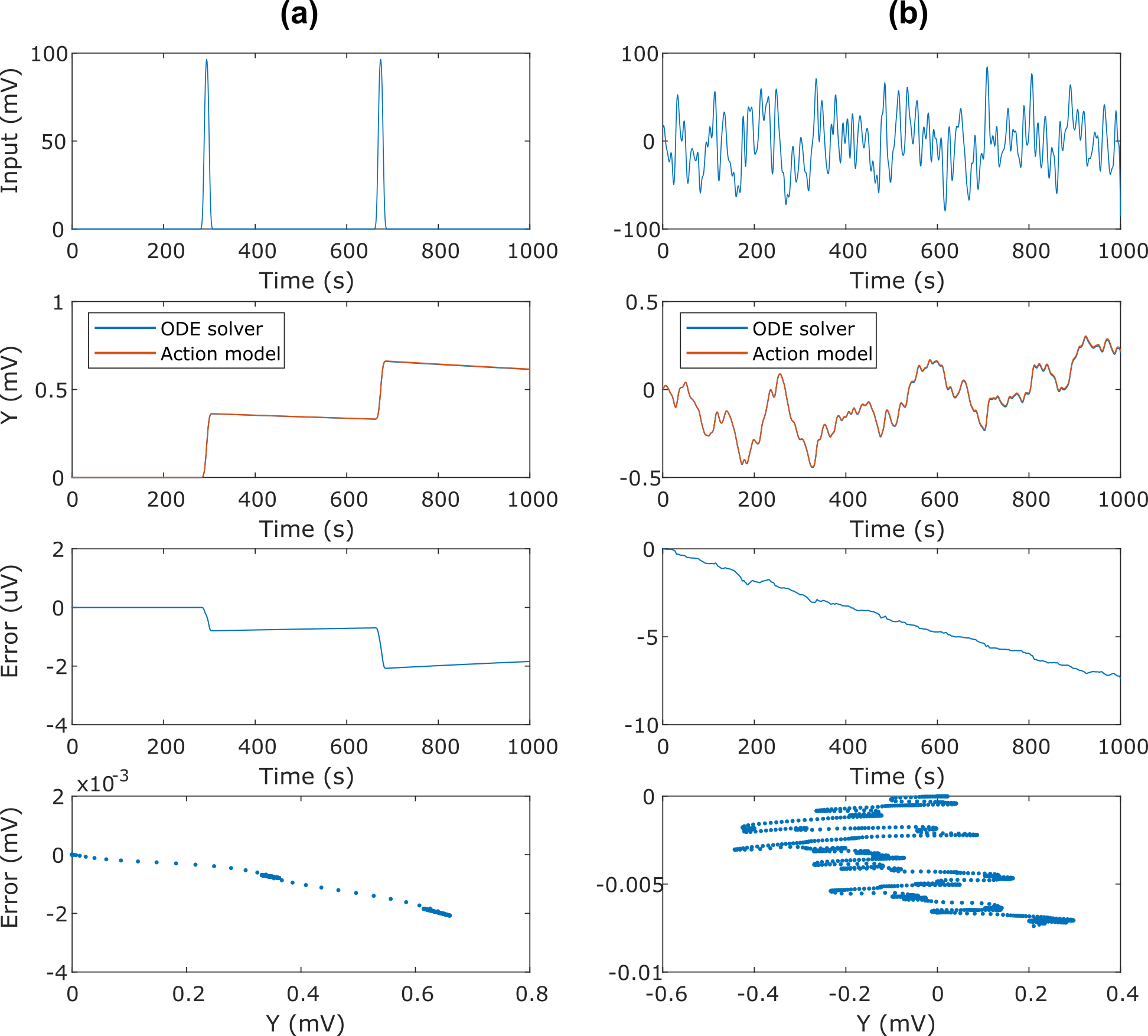}
		}
		\caption{Comparison between ODE solver and the action model for a) Short pulses and b) Small magnitude random signal. In either case, the action model closely tracks the ODE solver. The maximum error was less than \SI{10}{\micro\V} for both cases.}
		\label{modelCompare1}
	\end{figure}

\begin{figure}
		\centering
		{\includegraphics[width=7in]{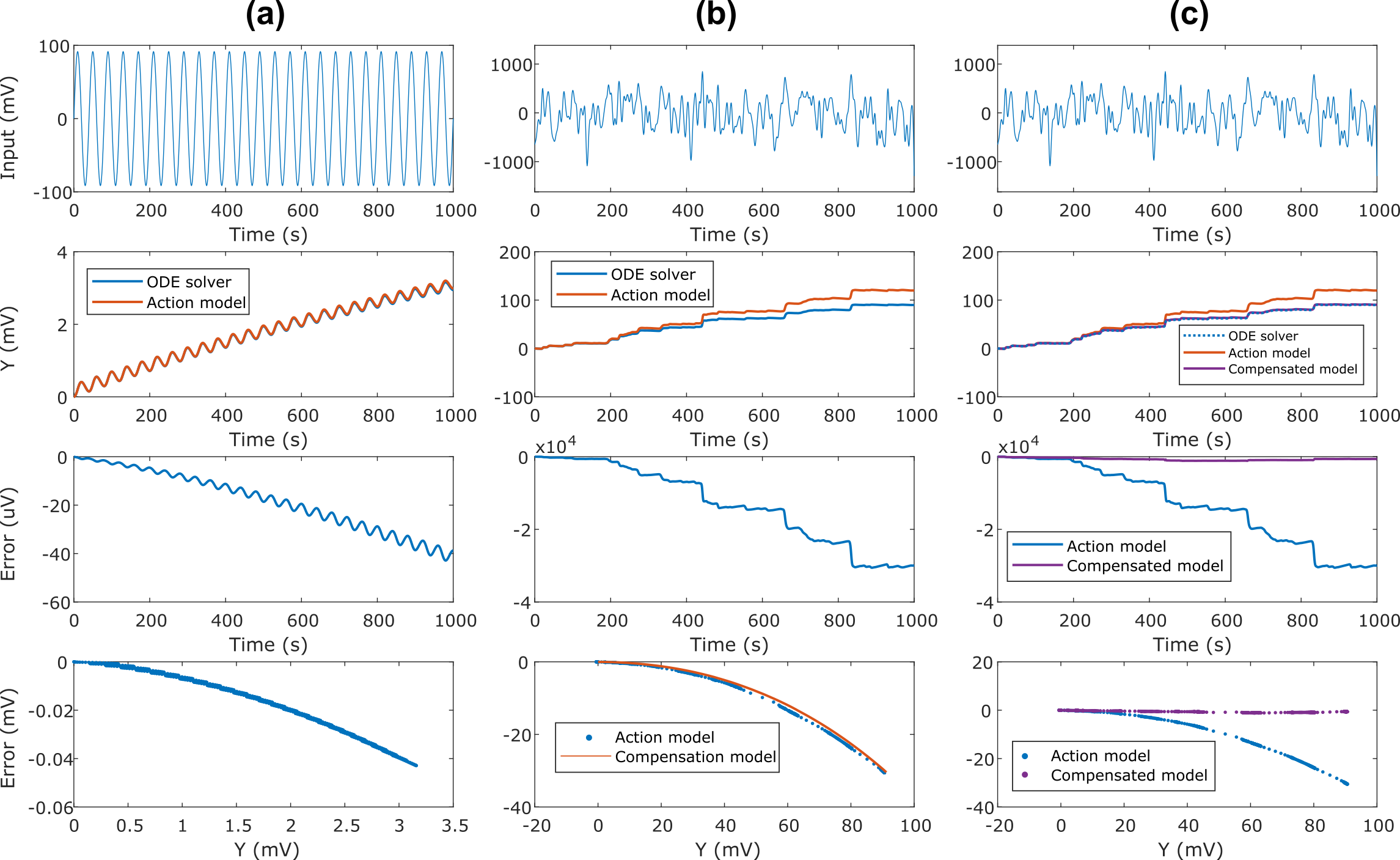}
		}
		\caption{Comparison between ODE solver and the action model a) Harmonic input signal leads to continuous increase in action, the error increases as action increases. b) Large continuous random signals lead to large errors in the action model compared to the ODE solver. However there is a relationship between the action and the error. An empirical model was fit that modeled error as a function of action. c) Compensation with this model leads to higher accuracy.  }
		\label{modelCompare2}
	\end{figure}

	The assumptions made in the derivation have been validated against a general ODE solver (Figs.~\ref{modelCompare1} and~\ref{modelCompare2}). As shown in the figure, the error was less than \SI{10}{\micro\V} for a response of \SI{1.5}{\mV}. Same analysis was run 100 times and the relative error was always less than $1\;\%$. The action model was computationally faster to solve than the ODE solver by a factor of $10^5$. The explicit action model led to large errors when the input signal was large (Fig.~\ref{modelCompare2}). The error arises due to assumption $Y{\ll}V$, made during linearizing the equation~16, to estimate the resynchronization of the response. For large $Y$, higher order terms can no longer be ignored and the resynchronization will be faster. As the 1st order model ignores these terms, it always overestimates the expected action at time T. The error in the model can be empirically reduced by fitting a model between the expected response (as generated by the ODE solver) and response calculated by the action model (Fig.~\ref{modelCompare2}c).
	
	\clearpage
	\subsection{AC analysis}
\begin{figure}[ht]
		\centering
		\includegraphics[width=7in]{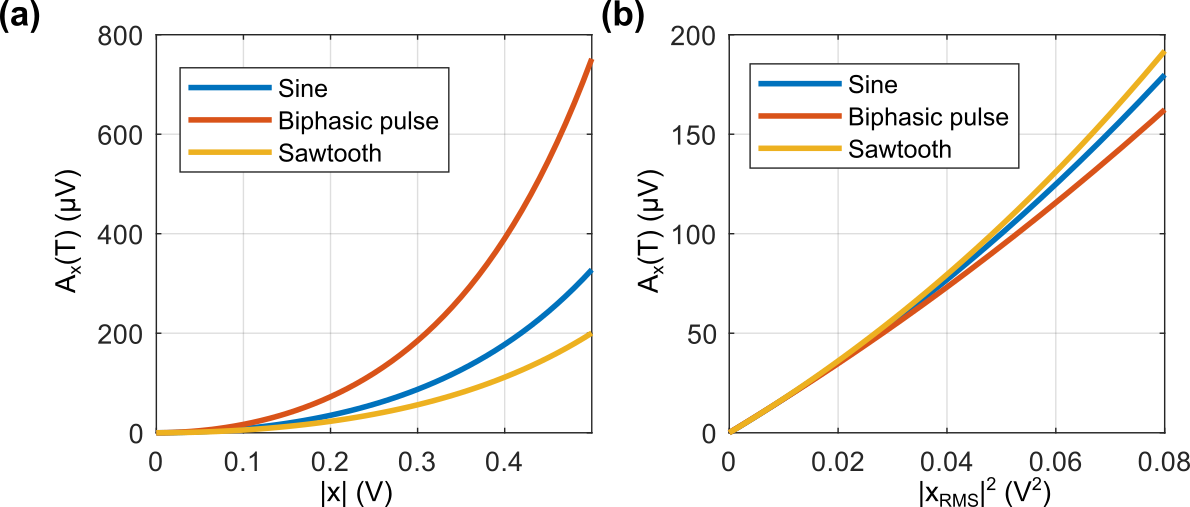}
		\caption{Simulation results for recorder response to AC signals. a) Action induced by signals of different shapes as a function of amplitude of the signal. System is sensitive to biphasic signals because of the rectification induced by FN tunneling. b) Action induced by signals of different shapes as a function of energy of the signal.}
		\label{AC analysis}
	\end{figure}

	The action induced by an AC coupled signal is monotonic with the energy of the signal. Actions due to different waveform shapes are more similar for signals with same energy (Fig.~\ref{AC analysis}b) compared to signals with same amplitude (Fig.~\ref{AC analysis}a). 

	\clearpage
	\subsection{Data retention model}
	Retention time for a given input signal was found by a fixed point method. First, a noise model was generated using experiments without any input modulation. Standard deviation ($\sigma_\m{t}$) was calculated across all runs as a function of time. Ideally, if the dynamics were perfectly synchronized, then the $\sigma_\m{t}$ obtained would be 0. However, we find that $\sigma_\m{t}$ increases with time due to integration of noise. We fit a rational equation on this noise. 
	\begin{equation*}
	\sigma_\m{t} = \frac{at}{t+b}
	\end{equation*}
	We chose this equation so that it stays bounded as time approaches $\infty$.
	Total noise in the system is given by
	\begin{equation*}
	N_\m{t} = \sigma_\m{t} + N_\m{0}
	\end{equation*}
	where $N_\m{0}$ is the noise associated with readout circuits and data acquisition system.
	
	The time of retention $T_\m{ret}$ was defined as the time instance at which the expected recorder response ($Y_\m{t}$ becomes lower than the predicted noise in the system $N_\m{t}$, i.e. the signal-to-noise ratio goes below unity. Thus, at time $t = T_\m{ret}$
	\begin{equation*}
	Y_{T_\m{ret}} = N_{T_\m{ret}}
	\end{equation*}
	
	\begin{figure}[ht]
		\centering
		{\includegraphics[width=7in]{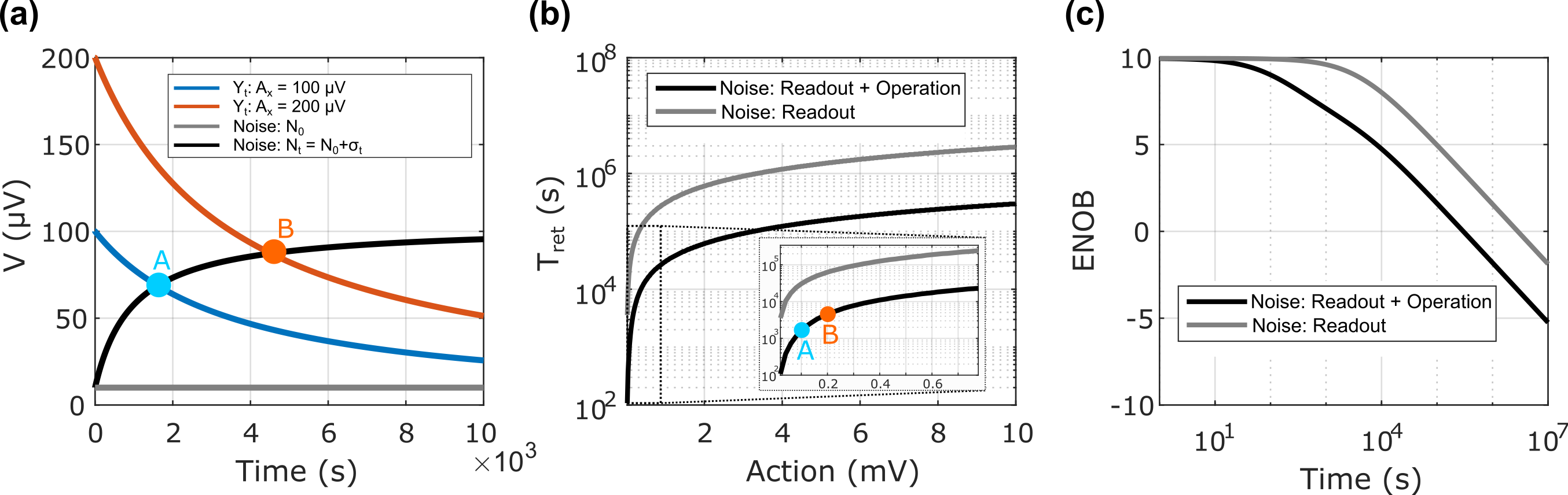}}
		\caption{a) Colored traces indicate recorder response for input signals with 100 and \SI{200}{\micro\V} actions. The gray trace indicates constant noise ($N_\m{0}$), due to readout and measurement circuits. Black curve indicates noise ($N_\m{t}$) due to unintended desynchronization occurring in absence of an input signal. It is obtained by adding the readout noise to the input referred noise ($\sigma_\m{t}$, assumed to be the standard deviation across trials for a recorder with no input signal - see variance in Fig. 2d). Intersection of the response curve with the noise curve (eg. Points A and B) is an estimation of system retention time, $T_\m{ret}$, at which point, the SNR of the system goes below 1. b) Retention time, $T_\m{ret}$, plotted as function of action for different noise profiles. Points A and B correspond to the intersection points in panel (a). $T_\m{ret}$ varies exponentially with action. c) Amount of information, measured as the effective number of bits (ENOB), stored in the system which decreases with time. }
		\label{fig_retention}
	\end{figure}	
	
	\clearpage
	\subsection{Parametric analysis}
\begin{figure}[ht]
		\centering
		\includegraphics[width=6.5in]{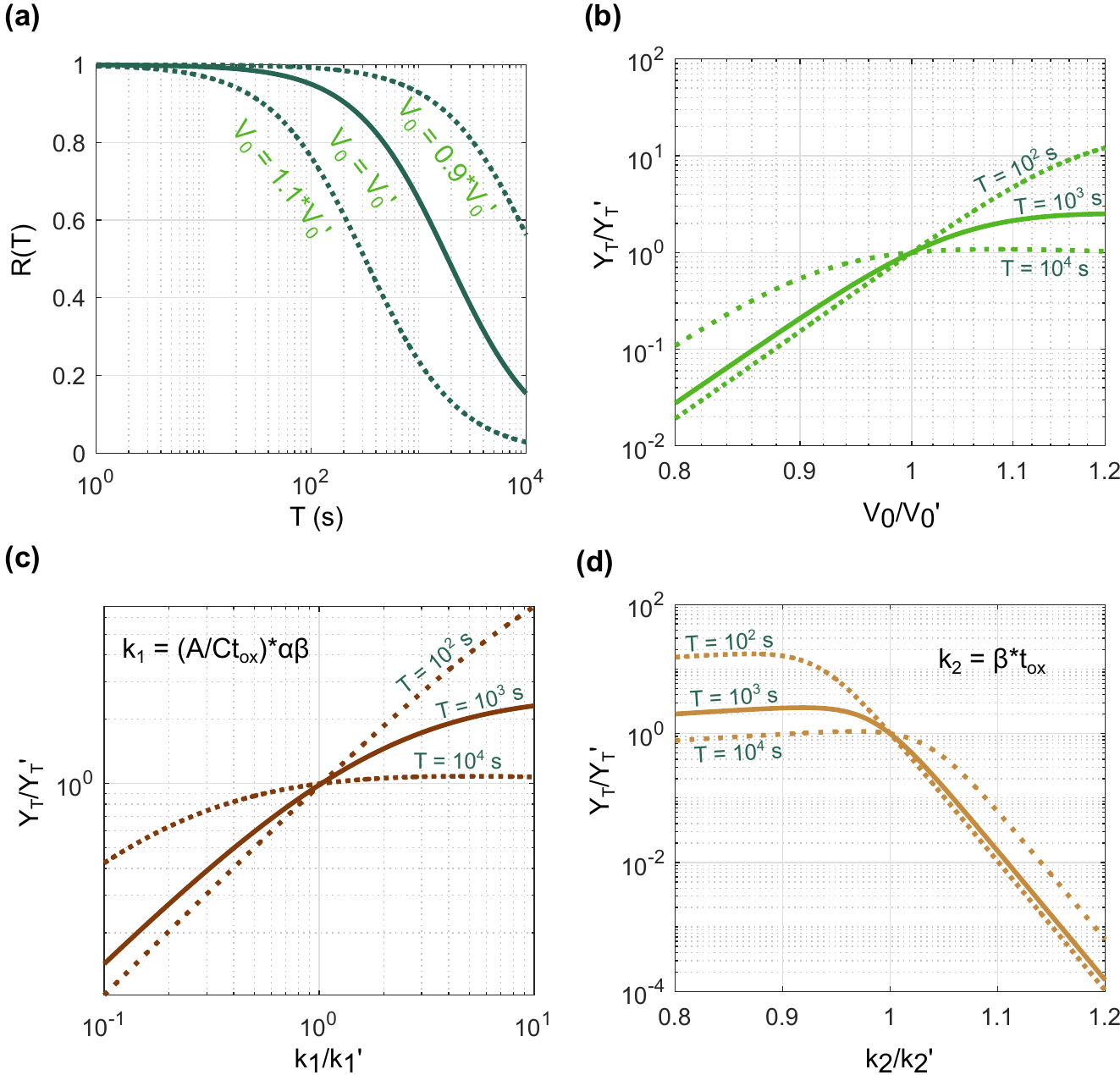}
		\caption{Simulation results for sensitivity and parametric analysis. Default parameters are $k_1 = \exp(38.5), k_2 = 346, V_0 = 7.5V $. $Y_\m{T}'$ is the baseline response at time $T$ for a single square pulse of magnitude \SI{100}{\mV} and duration 1 s. a) Time of sampling determines the amount of resynchronization b) Initial programming voltage affects the sensitivity of the recorder, but its effect becomes attenuated as time of sampling increases (due to resynchronization) c,d) Device parameters $k_1$ and $k_2$ can be tuned via system design and material selection to optimize the recorder response. $\alpha$ and $\beta$ are material parameters~\cite{lenzlinger1969fowler}.}
		\label{sysChar}
	\end{figure}

	Using modeling and simulations, we conducted parametric analysis for our system. Parameters $T$ (Fig.~\ref{sysChar}a) and $V_0$ ((Fig.~\ref{sysChar}b) are operational parameters that can be set at run time according to application requirements. $k_1$ (Fig.~\ref{sysChar}c) depends on the area of the tunneling junction and on the capacitance associated with the floating gate node. $k_1$ and $k_1$ (Fig.~\ref{sysChar}d) are also influenced by the thickness of the insulating material and other material properties like the barrier height at the interface between the conductor and the insulator.
	
	\clearpage
	\subsection{Impedance analysis and power estimation}\label{Power estimation}
\begin{figure}[ht]
		\centering
		\includegraphics[width=7in]{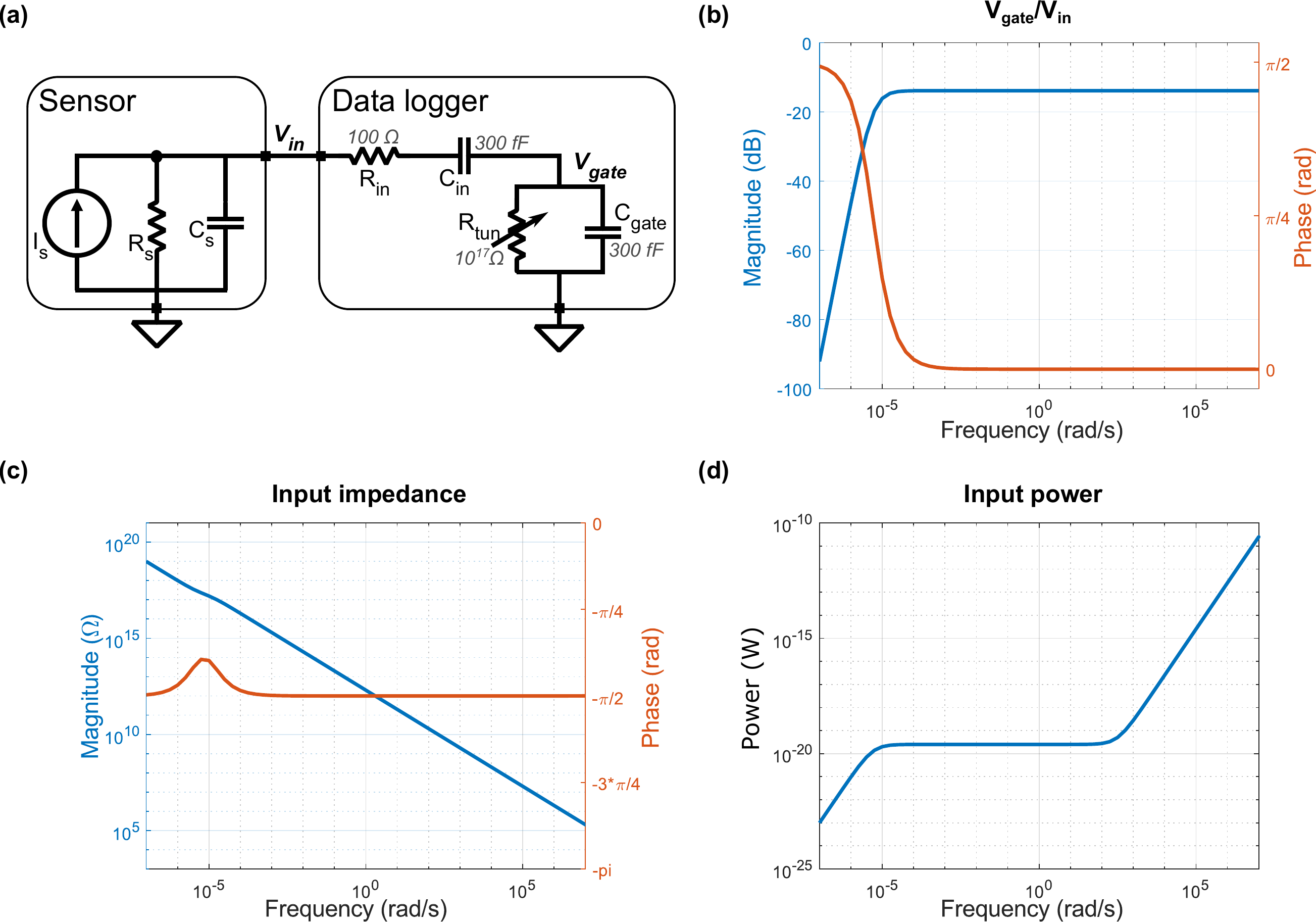}
		\caption{a) Equivalent circuit model of a sensor interfaced with sensor-data-logger. b) Relation of $V_\m{gate}$ to system input. c) Equivalent input impedance. d) Power spectral density for a matched system.}
		\label{sysImp}
	\end{figure}

	We performed simulation studies to characterize the input impedance of our sensor logger which can be modeled by the equivalent circuit shown in Fig.~\ref{sysImp}a. The DC input impedance of our system is on the order of \SI{{}e18}{\ohm}, since the input is connected to the gate of a MOSFET and the FN tunneling current is in the order of attoamperes. In this case, the impedance of the ESD protection diodes dominate the input impedance at DC frequency. We ignore this leakage path for our analysis and the high pass cut-off frequency is found to be at \SI{{}e-5}{\Hz} (Fig.~\ref{sysImp}b). At higher frequencies, the input capacitance and gate-to-substrate capacitance, along with input parasitic resistances create a low impedance path. However, this power is predominantly reactive in nature (Fig.~\ref{sysImp}c) and can be minimized with suitable source impedance matching. The power dissipated by the sensor-data-logger can be estimated as
	\begin{equation*}
	P(\omega) = \Re \left(\frac{V^2}{2Z_{in}(\omega)} \right)
	\end{equation*}
	Assuming that the natural dynamics of the FN device lie less than the frequency range of 1mHz, the average power dissipated can be estimated over the signal bandwidth of \SI{1}{\milli\Hz} to \SI{1}{\kilo\Hz} to be \SI{0.05}{\atto\watt}. So for an event lasting 100 seconds, energy dissipated is on the order of \SI{5}{\atto\joule}.
	
	\clearpage
	\subsection{Temporal dependence}
\begin{figure}[ht]
		\centering
		{\includegraphics[width=7in]{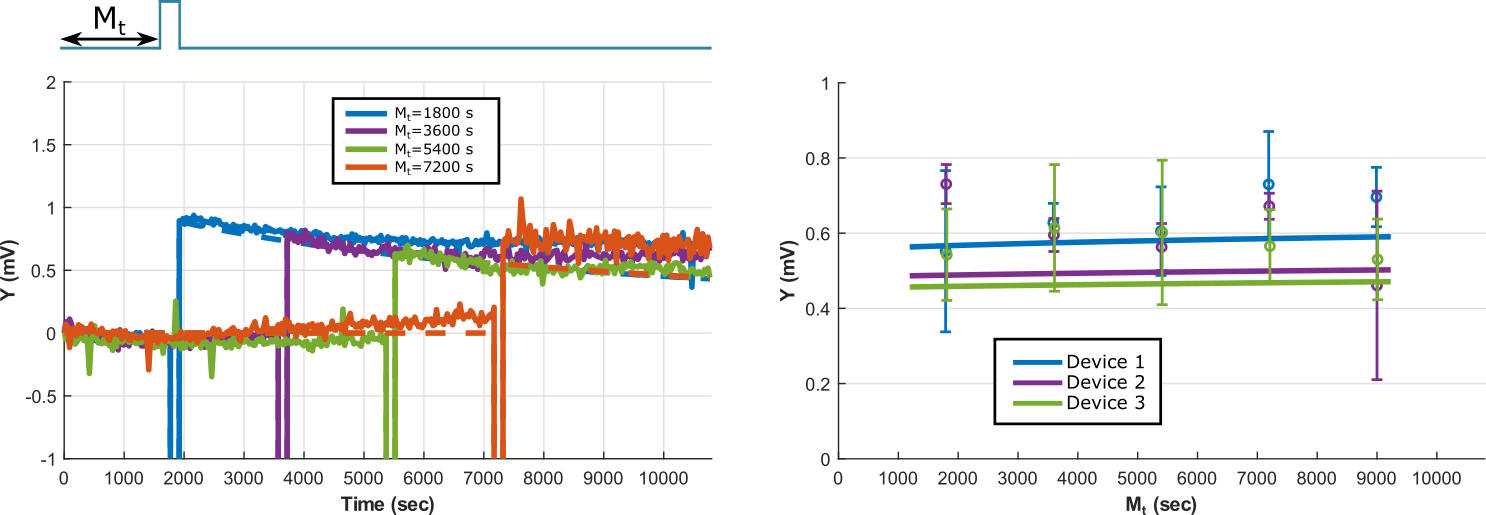}
		}
		\caption{Temporal dependence of recorder output to time of occurrence of an event. a) Recorder dynamics for events occurring at different times. b) Final output of three recorders, averaged over three trials, for events occurring at different times.}
		\label{YvsMt}
	\end{figure}

	Fig.~\ref{YvsMt} shows the weak dependence of the recorder output to the time of occurrence of events. Earlier events lead to larger desynchronization, but have more time to recover. Modeling studies show that the net result is that later events lead to a larger output at readout. However, the change in expected output was less than \SI{10}{\micro\V},  smaller than the errors arising due to measurement and operational desynchronization.
	\renewcommand\refname{Supplementary References}


\begin{thebibliography}{10}
		\expandafter\ifx\csname url\endcsname\relax
		\def\url#1{\texttt{#1}}\fi
		\expandafter\ifx\csname urlprefix\endcsname\relax\def\urlprefix{URL }\fi
		\providecommand{\bibinfo}[2]{#2}
		\providecommand{\eprint}[2][]{\url{#2}}
		
		\bibitem{indoorLight_mathews2014gaas}
		\bibinfo{author}{Mathews, I.}, \bibinfo{author}{Kelly, G.},
		\bibinfo{author}{King, P.~J.} \& \bibinfo{author}{Frizzell, R.}
		\newblock \bibinfo{title}{Gaas solar cells for indoor light harvesting}.
		\newblock In \emph{\bibinfo{booktitle}{2014 IEEE 40th Photovoltaic Specialist
				Conference (PVSC)}}, \bibinfo{pages}{0510--0513}
		(\bibinfo{organization}{IEEE}, \bibinfo{year}{2014}).
		
		\bibitem{wang2006piezoelectric}
		\bibinfo{author}{Wang, Z.~L.} \& \bibinfo{author}{Song, J.}
		\newblock \bibinfo{title}{Piezoelectric nanogenerators based on zinc oxide
			nanowire arrays}.
		\newblock \emph{\bibinfo{journal}{Science}} \textbf{\bibinfo{volume}{312}},
		\bibinfo{pages}{242--246} (\bibinfo{year}{2006}).
		
		\bibitem{autonomous_torah2008self}
		\bibinfo{author}{Torah, R.} \emph{et~al.}
		\newblock \bibinfo{title}{Self-powered autonomous wireless sensor node using
			vibration energy harvesting}.
		\newblock \emph{\bibinfo{journal}{Measurement science and technology}}
		\textbf{\bibinfo{volume}{19}}, \bibinfo{pages}{125202}
		(\bibinfo{year}{2008}).
		
		\bibitem{materials_mcevoy2015}
		\bibinfo{author}{McEvoy, M.~A.} \& \bibinfo{author}{Correll, N.}
		\newblock \bibinfo{title}{Materials that couple sensing, actuation,
			computation, and communication}.
		\newblock \emph{\bibinfo{journal}{Science}} \textbf{\bibinfo{volume}{347}},
		\bibinfo{pages}{1261689} (\bibinfo{year}{2015}).
		
		\bibitem{huang2011asynchronous}
		\bibinfo{author}{Huang, C.} \& \bibinfo{author}{Chakrabartty, S.}
		\newblock \bibinfo{title}{An asynchronous analog self-powered cmos
			sensor-data-logger with a 13.56 mhz rf programming interface}.
		\newblock \emph{\bibinfo{journal}{IEEE Journal of Solid-State Circuits}}
		\textbf{\bibinfo{volume}{47}}, \bibinfo{pages}{476--489}
		(\bibinfo{year}{2011}).
		
		\bibitem{chakrabartty2010self}
		\bibinfo{author}{Chakrabartty, S.}
		\newblock \bibinfo{title}{Self-powered sensor} (\bibinfo{year}{2010}).
		\newblock \bibinfo{note}{US Patent 7,757,565}.
		
		\bibitem{liangBioCas}
		\bibinfo{author}{Zhou, L.}, \bibinfo{author}{Abraham, A.~C.},
		\bibinfo{author}{Tang, S.~Y.} \& \bibinfo{author}{Chakrabartty, S.}
		\newblock \bibinfo{title}{A 5 nw quasi-linear cmos hot-electron injector for
			self-powered monitoring of biomechanical strain variations}.
		\newblock \emph{\bibinfo{journal}{IEEE transactions on biomedical circuits and
				systems}} \textbf{\bibinfo{volume}{10}}, \bibinfo{pages}{1143--1151}
		(\bibinfo{year}{2016}).
		
		\bibitem{mram_aakerman2005toward}
		\bibinfo{author}{{\AA}kerman, J.}
		\newblock \bibinfo{title}{Toward a universal memory}.
		\newblock \emph{\bibinfo{journal}{Science}} \textbf{\bibinfo{volume}{308}},
		\bibinfo{pages}{508--510} (\bibinfo{year}{2005}).
		
		\bibitem{fram1998physics}
		\bibinfo{author}{Auciello, O.}, \bibinfo{author}{Scott, J.~F.},
		\bibinfo{author}{Ramesh, R.} \emph{et~al.}
		\newblock \bibinfo{title}{The physics of ferroelectric memories}.
		\newblock \emph{\bibinfo{journal}{Physics today}}
		\textbf{\bibinfo{volume}{51}}, \bibinfo{pages}{22--27}
		(\bibinfo{year}{1998}).
		
		\bibitem{memristor_chua1971}
		\bibinfo{author}{Chua, L.}
		\newblock \bibinfo{title}{Memristor-the missing circuit element}.
		\newblock \emph{\bibinfo{journal}{IEEE Transactions on circuit theory}}
		\textbf{\bibinfo{volume}{18}}, \bibinfo{pages}{507--519}
		(\bibinfo{year}{1971}).
		
		\bibitem{ramadass2011battery}
		\bibinfo{author}{Ramadass, Y.~K.} \& \bibinfo{author}{Chandrakasan, A.~P.}
		\newblock \bibinfo{title}{A battery-less thermoelectric energy harvesting
			interface circuit with 35 mv startup voltage}.
		\newblock \emph{\bibinfo{journal}{IEEE Journal of Solid-State Circuits}}
		\textbf{\bibinfo{volume}{46}}, \bibinfo{pages}{333--341}
		(\bibinfo{year}{2011}).
		
		\bibitem{mercier2012energy}
		\bibinfo{author}{Mercier, P.~P.}, \bibinfo{author}{Lysaght, A.~C.},
		\bibinfo{author}{Bandyopadhyay, S.}, \bibinfo{author}{Chandrakasan, A.~P.} \&
		\bibinfo{author}{Stankovic, K.~M.}
		\newblock \bibinfo{title}{Energy extraction from the biologic battery in the
			inner ear}.
		\newblock \emph{\bibinfo{journal}{Nature biotechnology}}
		\textbf{\bibinfo{volume}{30}}, \bibinfo{pages}{1240} (\bibinfo{year}{2012}).
		
		\bibitem{ti_bq25505}
		\bibinfo{organization}{Texas Instruments}.
		\newblock \emph{\bibinfo{title}{bq25505 - ultra low-power boost charger with
				battery management and autonomous power multiplexer for primary battery in
				energy harvester applications}} (\bibinfo{year}{2019}).
		\newblock \urlprefix\url{http://www.ti.com/lit/ds/symlink/bq25505.pdf}.
		\newblock \bibinfo{note}{SLUSBJ3F}.
		
		\bibitem{informationSingleNode_appeltant2011}
		\bibinfo{author}{Appeltant, L.} \emph{et~al.}
		\newblock \bibinfo{title}{Information processing using a single dynamical node
			as complex system}.
		\newblock \emph{\bibinfo{journal}{Nature communications}}
		\textbf{\bibinfo{volume}{2}}, \bibinfo{pages}{1--6} (\bibinfo{year}{2011}).
		
		\bibitem{dynamicalInfo_dambre2012}
		\bibinfo{author}{Dambre, J.}, \bibinfo{author}{Verstraeten, D.},
		\bibinfo{author}{Schrauwen, B.} \& \bibinfo{author}{Massar, S.}
		\newblock \bibinfo{title}{Information processing capacity of dynamical
			systems}.
		\newblock \emph{\bibinfo{journal}{Scientific reports}}
		\textbf{\bibinfo{volume}{2}}, \bibinfo{pages}{1--7} (\bibinfo{year}{2012}).
		
		\bibitem{memoryTraces_ganguli2008}
		\bibinfo{author}{Ganguli, S.}, \bibinfo{author}{Huh, D.} \&
		\bibinfo{author}{Sompolinsky, H.}
		\newblock \bibinfo{title}{Memory traces in dynamical systems}.
		\newblock \emph{\bibinfo{journal}{Proceedings of the National Academy of
				Sciences}} \textbf{\bibinfo{volume}{105}}, \bibinfo{pages}{18970--18975}
		(\bibinfo{year}{2008}).
		
		\bibitem{mehta2019differential}
		\bibinfo{author}{Mehta, D.}, \bibinfo{author}{Raman, B.} \&
		\bibinfo{author}{Chakrabartty, S.}
		\newblock \bibinfo{title}{Differential fowler-nordheim tunneling dynamical
			system for attojoule sensing and recording}.
		\newblock In \emph{\bibinfo{booktitle}{2019 IEEE International Symposium on
				Circuits and Systems (ISCAS)}}, \bibinfo{pages}{1--5}
		(\bibinfo{organization}{IEEE}, \bibinfo{year}{2019}).
		
		\bibitem{liangTEDTimer}
		\bibinfo{author}{Zhou, L.} \& \bibinfo{author}{Chakrabartty, S.}
		\newblock \bibinfo{title}{Self-powered timekeeping and synchronization using
			fowler--nordheim tunneling-based floating-gate integrators}.
		\newblock \emph{\bibinfo{journal}{IEEE Transactions on Electron Devices}}
		\textbf{\bibinfo{volume}{64}}, \bibinfo{pages}{1254--1260}
		(\bibinfo{year}{2017}).
		
		\bibitem{lenzlinger1969fowler}
		\bibinfo{author}{Lenzlinger, M.} \& \bibinfo{author}{Snow, E.}
		\newblock \bibinfo{title}{Fowler-nordheim tunneling into thermally grown sio2}.
		\newblock \emph{\bibinfo{journal}{Journal of Applied physics}}
		\textbf{\bibinfo{volume}{40}}, \bibinfo{pages}{278--283}
		(\bibinfo{year}{1969}).
		
		\bibitem{harrison2001cmos}
		\bibinfo{author}{Harrison, R.~R.}, \bibinfo{author}{Bragg, J.~A.},
		\bibinfo{author}{Hasler, P.}, \bibinfo{author}{Minch, B.~A.} \&
		\bibinfo{author}{Deweerth, S.~P.}
		\newblock \bibinfo{title}{A cmos programmable analog memory-cell array using
			floating-gate circuits}.
		\newblock \emph{\bibinfo{journal}{IEEE Transactions on Circuits and Systems II:
				Analog and Digital Signal Processing}} \textbf{\bibinfo{volume}{48}},
		\bibinfo{pages}{4--11} (\bibinfo{year}{2001}).
		
		\bibitem{roundy2005effectiveness}
		\bibinfo{author}{Roundy, S.}
		\newblock \bibinfo{title}{On the effectiveness of vibration-based energy
			harvesting}.
		\newblock \emph{\bibinfo{journal}{Journal of intelligent material systems and
				structures}} \textbf{\bibinfo{volume}{16}}, \bibinfo{pages}{809--823}
		(\bibinfo{year}{2005}).
		
		\bibitem{siSensor_middelhoek2000celebration}
		\bibinfo{author}{Middelhoek, S.}
		\newblock \bibinfo{title}{Celebration of the tenth transducers conference: The
			past, present and future of transducer research and development}.
		\newblock \emph{\bibinfo{journal}{Sensors and Actuators A: Physical}}
		\textbf{\bibinfo{volume}{82}}, \bibinfo{pages}{2--23} (\bibinfo{year}{2000}).
		
		\bibitem{isfet_bergveld2003thirty}
		\bibinfo{author}{Bergveld, P.}
		\newblock \bibinfo{title}{Thirty years of isfetology: What happened in the past
			30 years and what may happen in the next 30 years}.
		\newblock \emph{\bibinfo{journal}{Sensors and Actuators B: Chemical}}
		\textbf{\bibinfo{volume}{88}}, \bibinfo{pages}{1--20} (\bibinfo{year}{2003}).
		
		\bibitem{ofet_torsi2013organic}
		\bibinfo{author}{Torsi, L.}, \bibinfo{author}{Magliulo, M.},
		\bibinfo{author}{Manoli, K.} \& \bibinfo{author}{Palazzo, G.}
		\newblock \bibinfo{title}{Organic field-effect transistor sensors: a tutorial
			review}.
		\newblock \emph{\bibinfo{journal}{Chemical Society Reviews}}
		\textbf{\bibinfo{volume}{42}}, \bibinfo{pages}{8612--8628}
		(\bibinfo{year}{2013}).
		
		\bibitem{smart_dust_warneke2001smart}
		\bibinfo{author}{Warneke, B.}, \bibinfo{author}{Last, M.},
		\bibinfo{author}{Liebowitz, B.} \& \bibinfo{author}{Pister, K.~S.}
		\newblock \bibinfo{title}{Smart dust: Communicating with a cubic-millimeter
			computer}.
		\newblock \emph{\bibinfo{journal}{Computer}} \textbf{\bibinfo{volume}{34}},
		\bibinfo{pages}{44--51} (\bibinfo{year}{2001}).
		
		\bibitem{neuralDust_seo2016wireless}
		\bibinfo{author}{Seo, D.} \emph{et~al.}
		\newblock \bibinfo{title}{Wireless recording in the peripheral nervous system
			with ultrasonic neural dust}.
		\newblock \emph{\bibinfo{journal}{Neuron}} \textbf{\bibinfo{volume}{91}},
		\bibinfo{pages}{529--539} (\bibinfo{year}{2016}).
		
		\bibitem{mehta_chakrabartty_2020}
		\bibinfo{author}{Mehta, D.} \& \bibinfo{author}{Chakrabartty, S.}
		\newblock \bibinfo{title}{Self-powered analog sensor-data-logger experimental
			data}  (\bibinfo{year}{2020}).
		\newblock \urlprefix\url{https://doi.org/10.6084/m9.figshare.12814592.v1}.
		
	\end{thebibliography}

\begin{thebibliography}{10}
		\expandafter\ifx\csname url\endcsname\relax
		\def\url#1{\texttt{#1}}\fi
		\expandafter\ifx\csname urlprefix\endcsname\relax\def\urlprefix{URL }\fi
		\providecommand{\bibinfo}[2]{#2}
		\providecommand{\eprint}[2][]{\url{#2}}
		
		\bibitem{lenzlinger1969fowler}
		\bibinfo{author}{Lenzlinger, M.} \& \bibinfo{author}{Snow, E.}
		\newblock \bibinfo{title}{Fowler-nordheim tunneling into thermally grown sio2}.
		\newblock \emph{\bibinfo{journal}{Journal of Applied physics}}
		\textbf{\bibinfo{volume}{40}}, \bibinfo{pages}{278--283}
		(\bibinfo{year}{1969}).
	\end{thebibliography}
\end{document}